\documentclass[twocolumn,english,pra]{revtex4}
\usepackage[T1]{fontenc}
\usepackage[latin9]{inputenc}
\usepackage{color}
\usepackage{babel}
\usepackage{amsmath}
\usepackage{graphicx}
\usepackage[unicode=true,pdfusetitle,
 bookmarks=true,bookmarksnumbered=false,bookmarksopen=false,
 breaklinks=false,pdfborder={0 0 1},backref=false,colorlinks=false]
 {hyperref}

\makeatletter
\@ifundefined{textcolor}{}
{%
 \definecolor{BLACK}{gray}{0}
 \definecolor{WHITE}{gray}{1}
 \definecolor{RED}{rgb}{1,0,0}
 \definecolor{GREEN}{rgb}{0,1,0}
 \definecolor{BLUE}{rgb}{0,0,1}
 \definecolor{CYAN}{cmyk}{1,0,0,0}
 \definecolor{MAGENTA}{cmyk}{0,1,0,0}
 \definecolor{YELLOW}{cmyk}{0,0,1,0}
}

\usepackage{multirow}

\makeatother

\begin{document}
\title{Slow semiclassical dynamics of a two-dimensional Hubbard model \\in
disorder-free potentials}
\author{Aleksander Kaczmarek, Adam S. Sajna}
\email{adam.sajna@pwr.edu.pl}
\affiliation{Department of Theoretical Physics, Faculty of Fundamental Problems
of Technology, Wroc\l{}aw University of Science and Technology, 50-370
Wroc\l{}aw, Poland}

\selectlanguage{english}%
\begin{abstract}
The quench dynamics of the Hubbard
model in tilted and harmonic potentials is discussed within the semiclassical picture. Applying the
fermionic truncated Wigner approximation (fTWA), the~dynamics of imbalances
for charge and spin degrees of freedom is analyzed and its time evolution
is compared with the exact simulations in one-dimensional lattice. Quench from
charge or spin density wave is considered. We show that introduction of harmonic and spin-dependent linear potentials sufficiently validates
fTWA for longer times. Such an improvement
of fTWA is also obtained for the higher order correlations in terms
of quantum Fisher information for charge and spin channels. This allows
us to discuss the dynamics of larger system sizes and connect
our discussion to the recently introduced Stark many-body localization.
In particular, we focus on a finite two-dimensional system and show
that at intermediate linear potential strength, the addition of a harmonic potential and spin dependence of the tilt, results in subdiffusive
dynamics, similar to that of disordered systems. Moreover, for specific
values of harmonic potential, we observed phase separation of ergodic
and non-ergodic regions in real space. The latter fact is especially
important for ultracold atom experiments in which harmonic confinement
can be easily imposed, causing a significant change in~relaxation
times for different lattice locations.
\end{abstract}
\maketitle

\section{Introduction}

The search for robust quantum many-body systems which show no thermalization
or whose thermalization is very slow, has become a focus of a number
of theoretical and experimental investigations (see e.g. \citep{RevModPhys.91.021001,Nandkishore2015,vanHorssen2015,Gogolin_2016,PhysRevLett.118.266601,Grover_2014,PhysRevLett.119.176601,PhysRevB.101.174204,Scherg2021,PhysRevX.10.011047}
and references therein). The best known example in closed systems
that show robust non-ergodic behavior is the many-body localized (MBL)
phase \citep{PhysRevLett.95.206603,Basko2006,PhysRevB.75.155111,Luitz2017}.
MBL systems are considered as potential models for quantum memory
devices \citep{Nandkishore2015,PhysRevLett.113.147204} and are relevant
for quantum computational problems \citep{Laumann2015}. MBL behavior
comes from the interplay of a disorder and interactions and such systems
have already been realized experimentally on many platforms like ultracold
atoms in optical lattices, trapped ions and superconducting qubits
\citep{Smith2016,PhysRevResearch.4.013148,mblschreiber,Choi2016,Lukin256}.
However, it has been recently shown that MBL features can also be
observed in the systems without quenched disorder but showing a linear
and weak harmonic potential \citep{PhysRevLett.122.040606}. Another
possibility is to add a weak disorder potential to a tilted lattice
\citep{vanNieuwenburg2019}. Such a phenomenon has been named the
Stark many-body localization (SMBL) and some of its features have
already been investigated experimentally \citep{Smith2019,PhysRevLett.127.240502,Morong2021-mh,PhysRevX.10.011042}.

Focusing on the one-dimensional dynamical behavior of SMBL we have
to mention the non-decaying character of the imbalance function \citep{PhysRevLett.122.040606,vanNieuwenburg2019,PhysRevB.102.054206},
the appearance of logarithmic-in-time growth of entanglement entropy,
quantum Fisher information (QFI) and quantum mutual information \citep{PhysRevLett.122.040606,PhysRevB.102.104203,PhysRevResearch.2.032039,Yao2021,PhysRevB.104.014201,PhysRevLett.127.240502,PhysRevB.102.054206},
non-ergodic behavior of the squared width of the excitation~\cite{PhysRevB.105.L140201} and average participation ratio which is directly
related to the return probability \citep{vanNieuwenburg2019}. For
two-dimensional systems, much less is known about a possible SMBL
behavior. It seems that the absence of rare regions can lead to non-ergodic behavior in the thermodynamic limit
\citep{vanNieuwenburg2019}. However, strongly non-ergodic polarized
regions \citep{PhysRevB.103.L100202}, which can lead to the SMBL
phase in the thermodynamic limit of one-dimensional systems, are less
relevant in two dimensions. Therefore the existence of SMBL in higher
dimensional systems can be questioned \citep{PhysRevB.105.134204}. This conclusion is consistent with the experimental observation that the presence of defects in polarized regions can lead to subdiffusive behavior~\cite{PhysRevX.10.011042}.
Moreover, going beyond the linear potential e.g. by adding harmonicity
to the lattice, can lead to various dynamical types of behavior depending
on the lattice location. Such an analysis, for one-dimensional systems,
has recently been given in the context of SMBL \citep{Yao2021,PhysRevResearch.2.032039,PhysRevB.102.104203}
leaving two-dimensional systems unexplored.

In this work, we focus on the disorder-free quantum evolution of the weakly polarized initial states and point out dynamical similarities with disordered systems in one and two dimensions. We give an approximate description of the quench dynamics
from density waves with a short wavelength which evolve under a wide range of tilt strength (density waves with a short wavelength correspond to the weakly polarized initial states which can be more easily delocalized~\cite{PhysRevB.105.134204}). In contrast to the recent studies of quantum dynamics in two dimensions \citep{PhysRevX.10.011042,PhysRevB.105.134204} we mostly assume that the field gradient is applied at an irrational
angle in order to remove the equipotential directions \citep{vanNieuwenburg2019}.
In particular, we show that a finite two-dimensional lattice system
with relatively weak harmonic potential and sufficiently strong tilt
exhibits subdiffusive dynamical behavior similar to that known for
disordered systems \citep{BarLev2016,PhysRevA.102.033338}. We achieve
this by analyzing the quantum dynamics of the Hubbard model which
can be directly experimentally realized \citep{Scherg2021,mblschreiber,PhysRevLett.116.140401,PhysRevLett.119.260401,PhysRevX.7.041047,PhysRevLett.122.170403,PhysRevX.10.011042}.
In our numerical study, we exploit fermionic truncated Wigner approximation
(fTWA) to deal with system of larger sizes \citep{Davidson2017,S.M.Davidson.thesis,PhysRevA.102.033338,PhysRevB.99.134301,2007.05063,2205.06620}.
Such an analysis is possible because fTWA gives a reliable description
in the parameter space in which together with the tilt potential, a harmonic
potential has been added to the lattice and a spin dependence of the linear field has been taken into account. The importance of the spin-dependent local potential has been previously
linked to the full MBL in the disordered Hubbard system because it is responsible for the localization of the spin
degrees of freedom \citep{PhysRevB.94.241104}. Here we observe a
similar effect for spin dynamics on a tilted lattice and demonstrate
that the prediction of fTWA dynamics is highly enhanced in this limit.

To discuss the dynamics of a Hubbard model on the tilted lattice we
focus our analysis on the imbalance and QFI for charges and spins.
Both observables are related to the on-site density measurements and
are experimentally accessible \citep{mblschreiber,Choi2016,PhysRevLett.116.140401,PhysRevX.7.041047,Lschen2017,Smith2016,Smith2019}.
Imbalance and QFI were chosen because both are well-established indicators
of non-ergodicity. Moreover QFI can distinguish the Wannier-Stark
localization from SMBL through a logarithmic-in-time type growth in
the SMBL phase \citep{PhysRevLett.127.240502}. In this work, we show
that in two dimensions QFI exhibits a slow logarithmic-like growth
which is similar to the QFI behavior of disordered systems \citep{Smith2016,PhysRevB.99.054204,PhysRevB.99.241114,Guo2020,PhysRevA.102.033338}
and recently studied tilted triangular ladder \citep{PhysRevLett.127.240502}.
Moreover, we discuss the way in which harmonic potential together
with spin-dependent tilt causes a change in the charge imbalance
decay from diffusive to subdiffusive behavior for intermediate strength
of linear potential. Interestingly for spins we show that the decay
of imbalance is even more pronounced and changes from superdiffusive
to subdiffusive behavior. It is worth stressing that due to the approximation
made in studying dynamical behavior, we cannot conclude about a possibility
of a transition to SMBL phase in two dimensions. However, we can indicate
certain dynamical features which are difficult to handle by other
computational methods.

Finally, the fTWA method also enables us to discuss the appearance of phase
separation of ergodic and non-ergodic long-lived phases in a two-dimensional
lattice, which is an extension of previous theoretical studies performed
for one-dimensional lattices \citep{Yao2021,PhysRevResearch.2.032039,PhysRevB.102.104203}.

The manuscript is constructed as follows. In Sec. \ref{sec: fTWA-for-Hubbard},
the fTWA method is shortly discussed. In Sec. \ref{sec: benchmark},
the benchmark of fTWA method against exact diagonalization (ED) in
one-dimensional Hubbard system is provided together with the mean
square error analysis (MSE) for imbalances and QFI. It is realized
for the charge and spin density wave initial conditions and the roles
of harmonic and spin-dependent linear potentials are described. The
two-dimensional analysis of the many-body dynamics in tilted lattices
is given in Sec. \ref{sec: 2D}. The paper ends with a summary of the
obtained results (Sec. \ref{sec: Summary}).

\section{fTWA for the Hubbard model in disordered-free potentials \label{sec: fTWA-for-Hubbard}}

Before we define the semiclassical dynamics within fTWA we begin with writing
the Hubbard Hamiltonian in terms of the creation $\hat{c}_{i\sigma}^{\dagger}$
and annihilation $\hat{c}_{i\sigma}$ operators
\begin{equation}
H=-\sum_{ij,\,\sigma}J_{ij}\hat{c}_{i\sigma}^{\dagger}\hat{c}_{j\sigma}+U\sum_{i}\hat{n}_{i\uparrow}\hat{n}_{i\downarrow}+\sum_{i,\sigma}\Delta(i,\sigma)\hat{n}_{i\sigma},
\end{equation}
where the operator $\hat{c}_{i\sigma}^{\dagger}$ ($\hat{c}_{i\sigma}$)
creates (annihilates) fermionic particle at position $i$ with spin
$\sigma\in\left\{ \uparrow,\,\downarrow\right\} $, $\hat{n}_{i\sigma}=\hat{c}_{i\sigma}^{\dagger}\hat{c}_{i\sigma}$
is the density operator, $J_{ij}$ is the hopping energy, $\Delta(i,\sigma)$
is the spin-dependent on-site potential and $U$ is the on-site interaction
energy between two spin species. Throughout this work it is assumed that
$J_{ij}$ is non-zero for the nearest neighbour sites only for which we set $J_{ij}=J$.
Then, instead of solving the Schr\"odinger equation, approximated
quantum dynamics in fTWA is obtained by equating Hamilton
equations of motion with the addition of quantum fluctuation encoded in
the initial conditions through the Wigner function $W$ \citep{Polkovnikov2010,Davidson2017}.
Equations of motion for the Hubbard take the form \citep{Davidson2017,PhysRevA.102.033338}
\begin{align}
 & i\frac{d\rho_{m\sigma n\sigma}}{dt}=-\sum_{k}\left(J_{nk}\rho_{m\sigma,k\sigma}-J_{km}\rho_{k\sigma,n\,\sigma}\right)\nonumber \\
 & +\rho_{m\sigma n\sigma}\left[\Delta(n,\sigma)-\Delta(m,\sigma)+U\left(\rho_{n-\sigma n-\sigma}-\rho_{m-\sigma m-\sigma}\right)\right],\label{eq: Hamilton eq}
\end{align}
where $\rho_{m\sigma n\sigma}$ are phase space variables corresponding
to fermionic bilinears $\hat{E}_{m\sigma}^{n\sigma}=\left(\hat{c}_{n\sigma}^{\dagger}\hat{c}_{m\sigma}-\hat{c}_{m\sigma}^{\dagger}\hat{c}_{n\sigma}\right)/2$
($\rho_{n\sigma m\sigma}$ are obtained by the Wigner-Weyl quantization
procedure \citep{Davidson2017}). Here the so-called $\rho$ representation
of Hamiltonian $H$ was used \citep{Davidson2017,PhysRevA.102.033338}.
In order to obtain the expectation value of a given observable, e.g. $\hat{\mathcal{O}}$,
trajectories are sampled from the initial Wigner function $W(\boldsymbol{\rho}_{0})$
and summed up according to the following procedure
\begin{equation}
\left\langle \hat{\mathcal{O}}(t)\right\rangle \overset{\text{fTWA}}{\approx}\int\mathcal{O}_{W}(\boldsymbol{\rho}(t))W(\boldsymbol{\rho}_{0})d\boldsymbol{\rho}_{0}=\left\langle \mathcal{O}_{W}(t)\right\rangle _{cl},\label{eq: expectation value in fTWA}
\end{equation}
where $\mathcal{O}_{W}$ is a Weyl symbol of $\hat{\mathcal{O}},$
$\boldsymbol{\rho}(t)=\left\{ \rho_{i\sigma j\sigma'}(t):\,i,j\in\left\{ 1,2,...,N\right\} ,\,\sigma,\,\sigma'\in\left\{ \uparrow,\,\downarrow\right\} \right\} $,
$N$ is the number of sites, $\boldsymbol{\rho}_{0}=\boldsymbol{\rho}(t=0)$.
Initial conditions encoded in the Wigner function $W(\boldsymbol{\rho}_{0})$
are obtained by approximating $W(\boldsymbol{\rho}_{0})$ as multivariate
Gaussians and reading off its first and second moments from matching the semiclassical
and quantum expectation values \citep{Davidson2017}.

Except for non-interacting systems, fTWA gives an accurate description
of general systems only in the early times \citep{Polkovnikov2010}.
However, in the next section, we numerically show that slight modification
of the linear potential leads to the improvement of the long-time fTWA predictions.
In one-dimensional systems, we consider the following form of the onsite
potential
\begin{equation}
\Delta(j,\sigma)=\Delta_{1}\left(\delta_{\sigma\downarrow}+A\delta_{\sigma\uparrow}\right)j+\Delta_{2}(j-j_{0})^{2},\label{eq: on-site potential}
\end{equation}
where $\Delta_{1}$ ($\Delta_{2}$) is the strength of linear (harmonic)
potential, $A$ introduce a spin dependence to the linear potential for any $A\neq1$. In this work a weak spin dependence 
($A=0.9$) is considered as in the recent experiment by S. Scherg et al. \citep{Scherg2021}. In Sec. \ref{sec: 2D}
we assume a two-dimensional system and then the potential
is modified correspondingly.

Throughout the paper, the interaction strength is set to $U/J=1$ and open
boundary conditions are assumed.

\section{The role of harmonic potential and spin dependence of the linear field \label{sec: benchmark}}

\begin{figure*}
\includegraphics[scale=0.38]{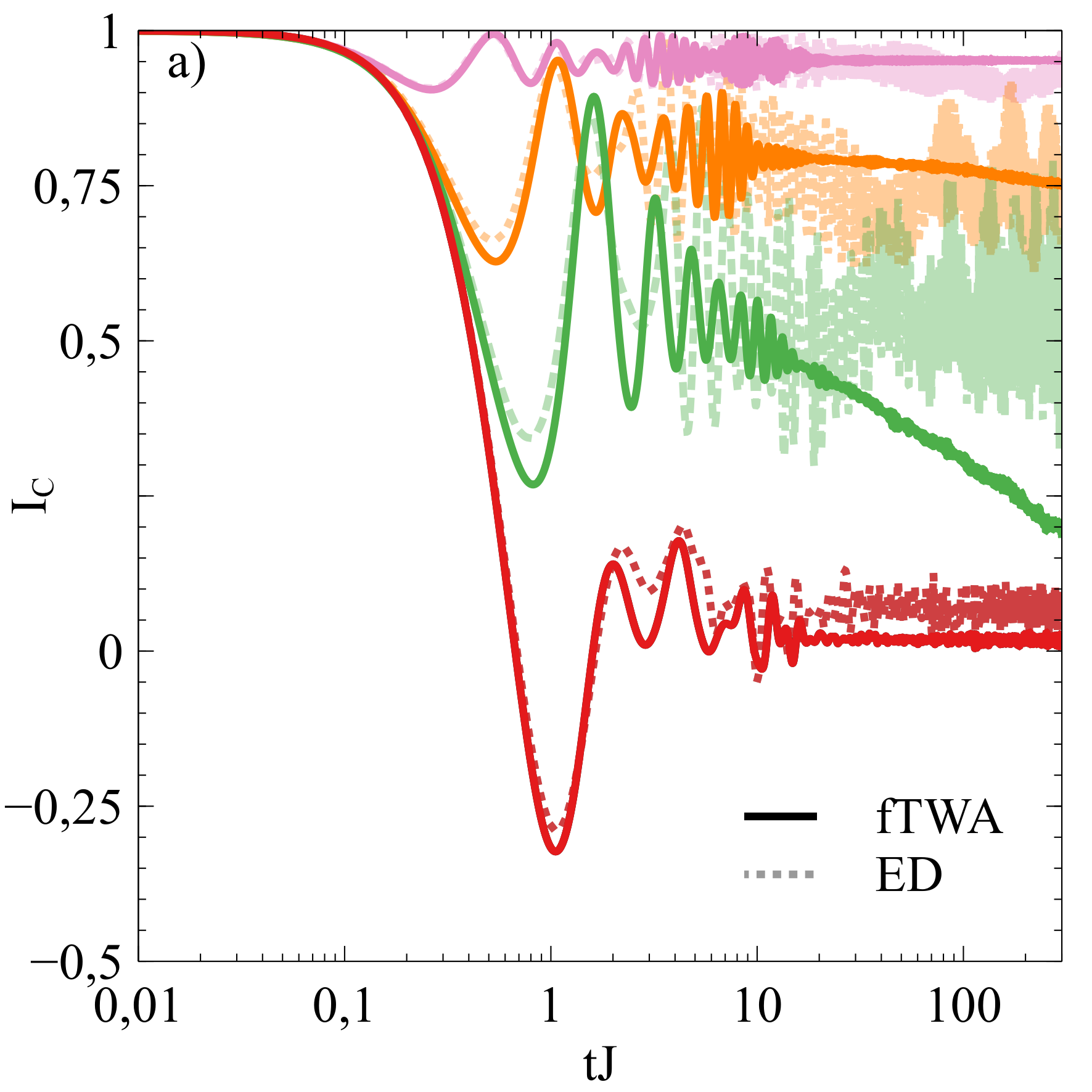}\includegraphics[scale=0.38]{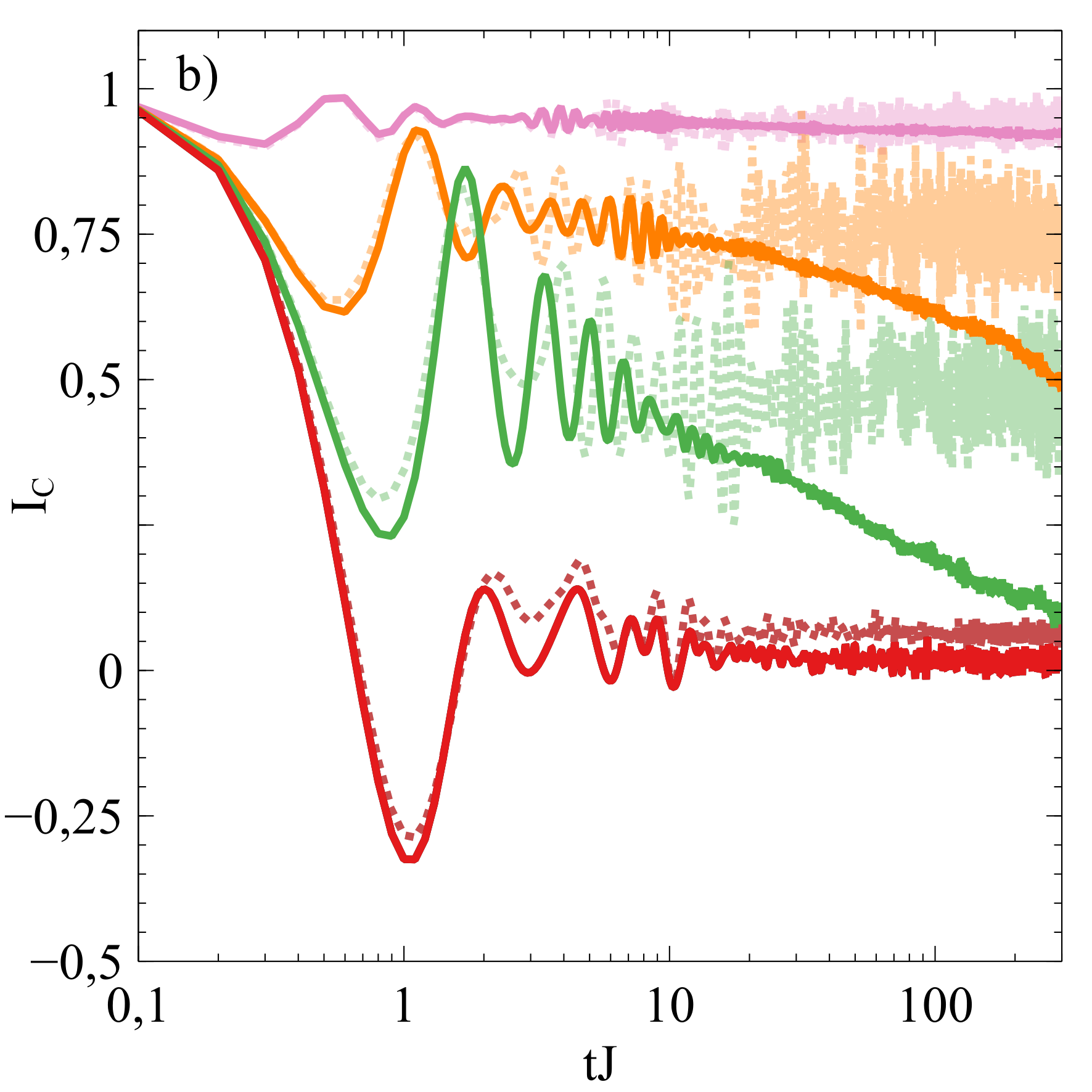}\includegraphics[scale=0.38]{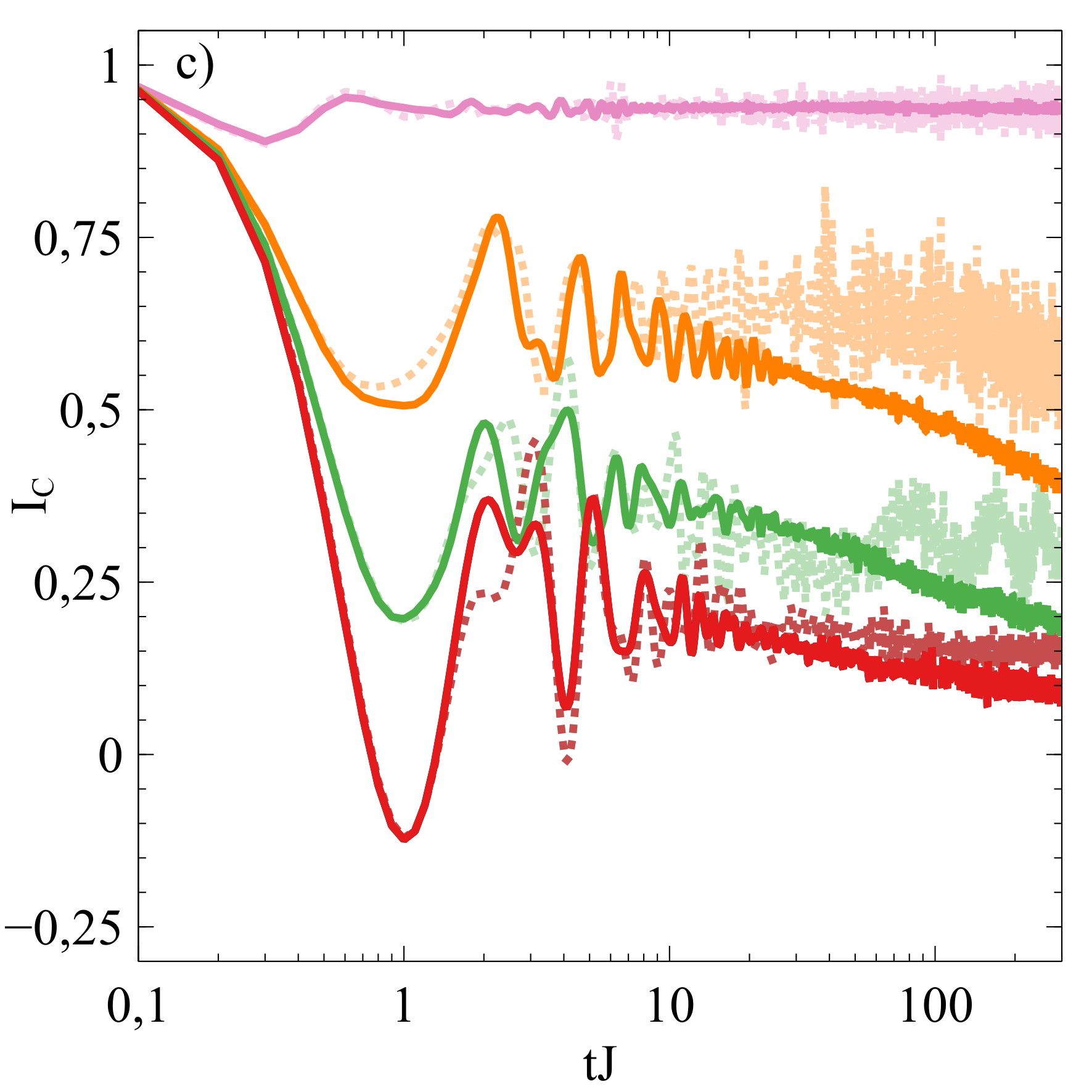}
\includegraphics[scale=0.38]{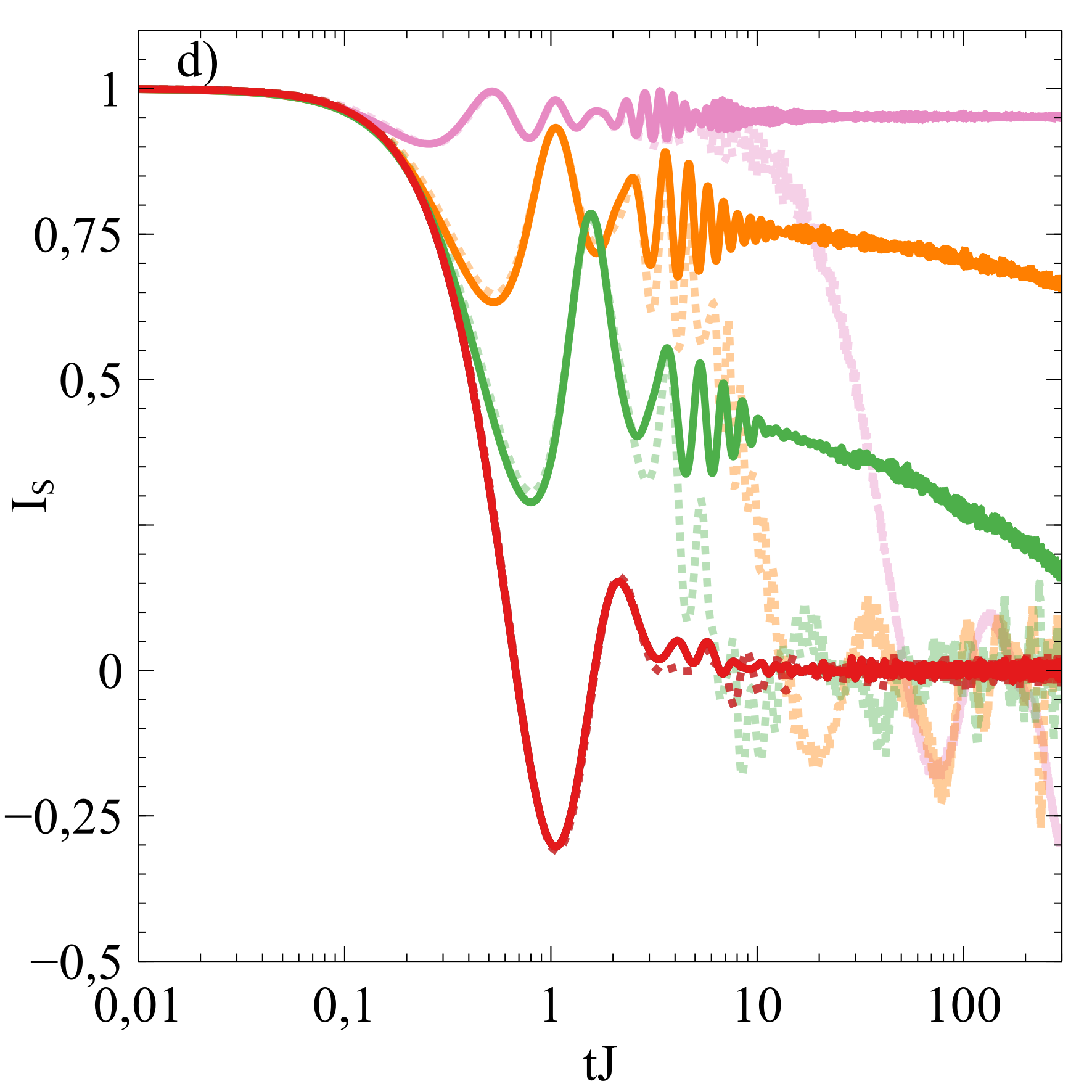}\includegraphics[scale=0.38]{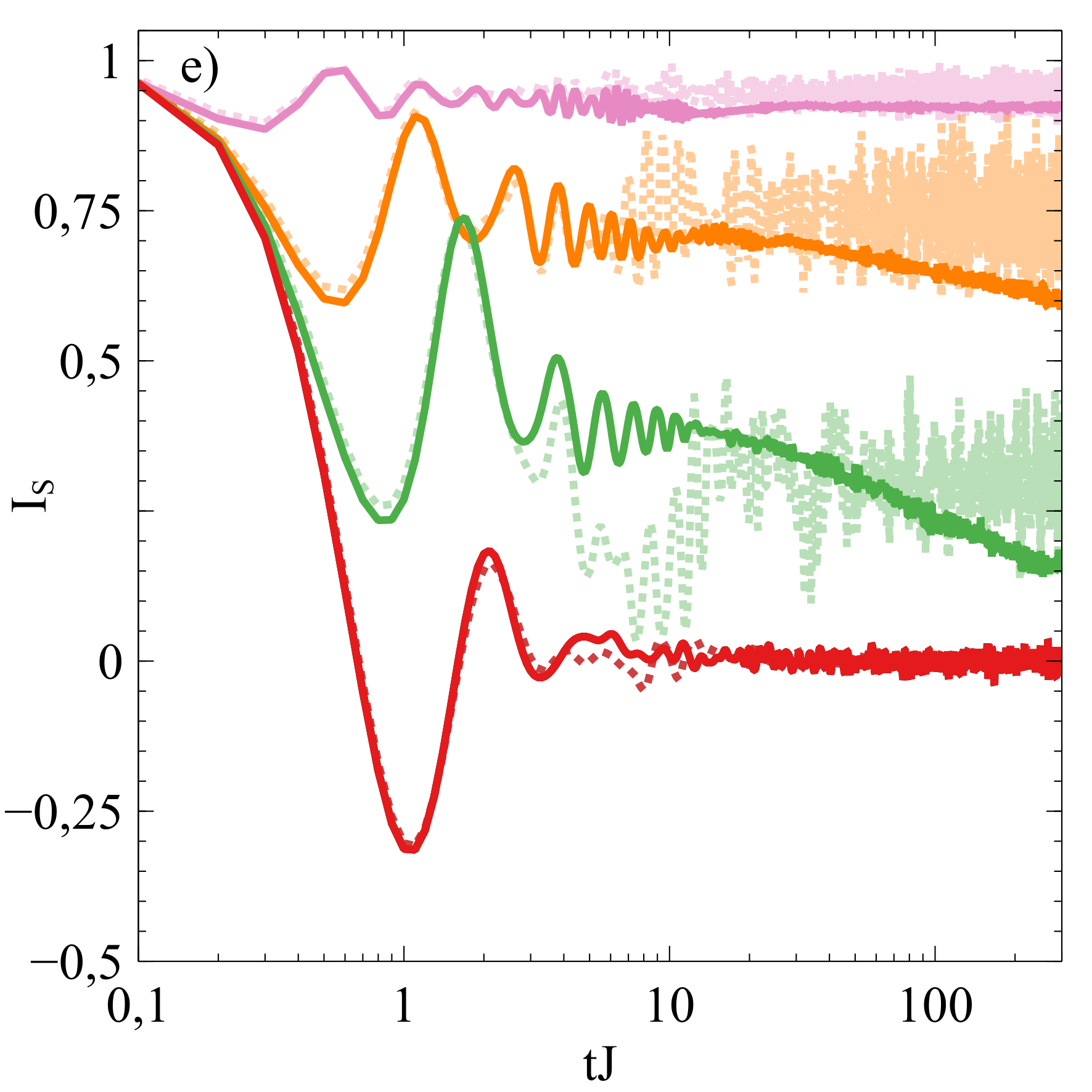}\includegraphics[scale=0.38]{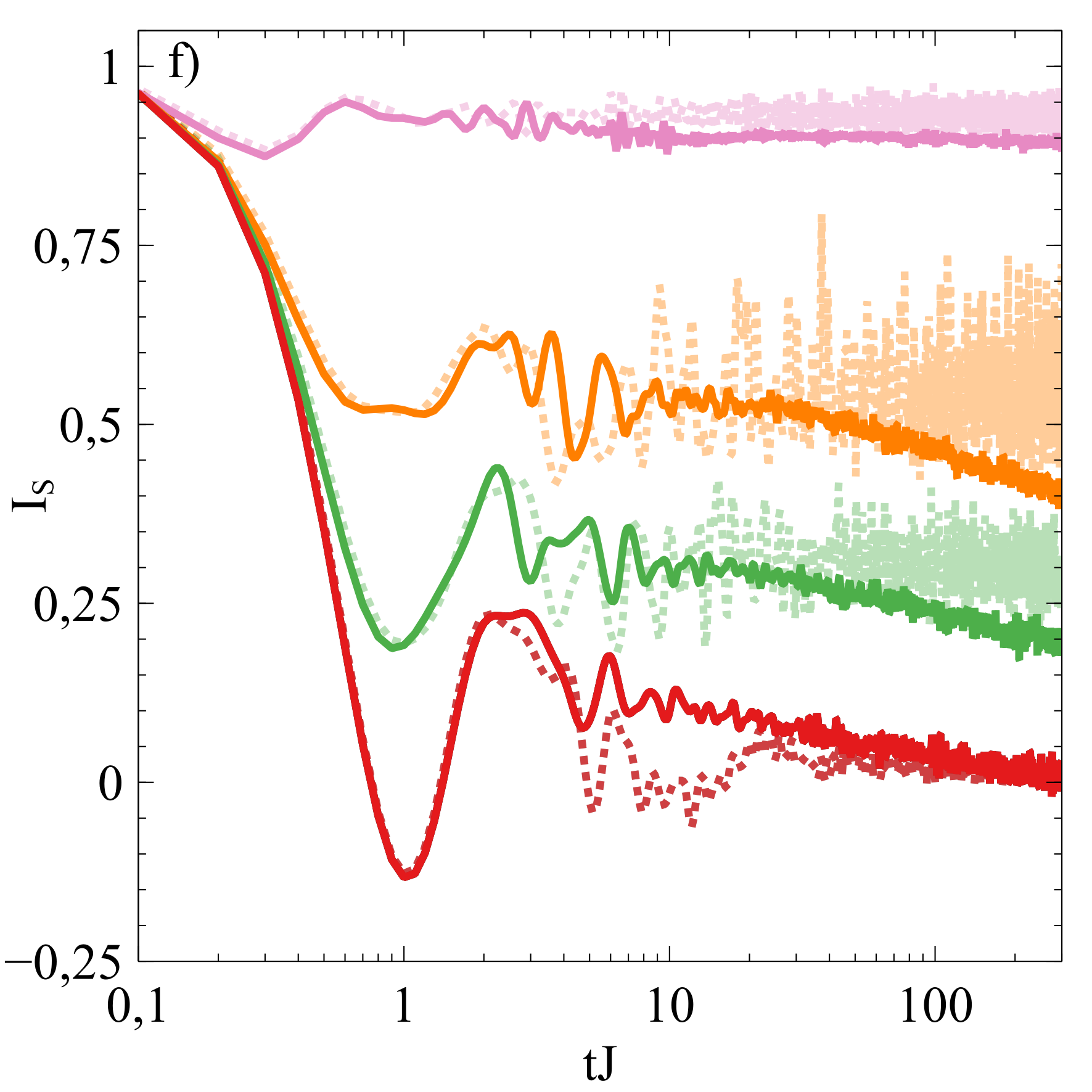}
\caption{Time dependences of imbalance functions for charges (Fig. a-c) and
spins (Fig. d-f). In each plot different strengths of the linear potential
$\Delta_{1}/J$ are taken, i.e. $\Delta_{1}/J=1,\,4,\,6,\,12$ from the bottom to top. The dashed lines indicate the fTWA, while
solid lines correspond to the ED results. The first column (a and d) corresponds to $A=1$,
$\Delta_{2}=0$, the second column (b and e) to $A=0.9$, $\Delta_{2}=0$, and the
third column (c and f) to $A=0.9$, $\Delta_{2}/J=0.5$. Simulations
are performed for the one-dimensional system with 8 sites and with
the CDW (a-c) or SDW (d-f) initial conditions. The other parameters
are $U/J=1$, $j_{0}=4$, the number of trajectories used in fTWA is 1000
or higher. Preliminary results of (a) were obtained in \citep{kaczmarek.thesis}.\textcolor{red}{\label{fig: imbalances time dependence}}}
\end{figure*}

\begin{figure*}
\includegraphics[scale=0.45]{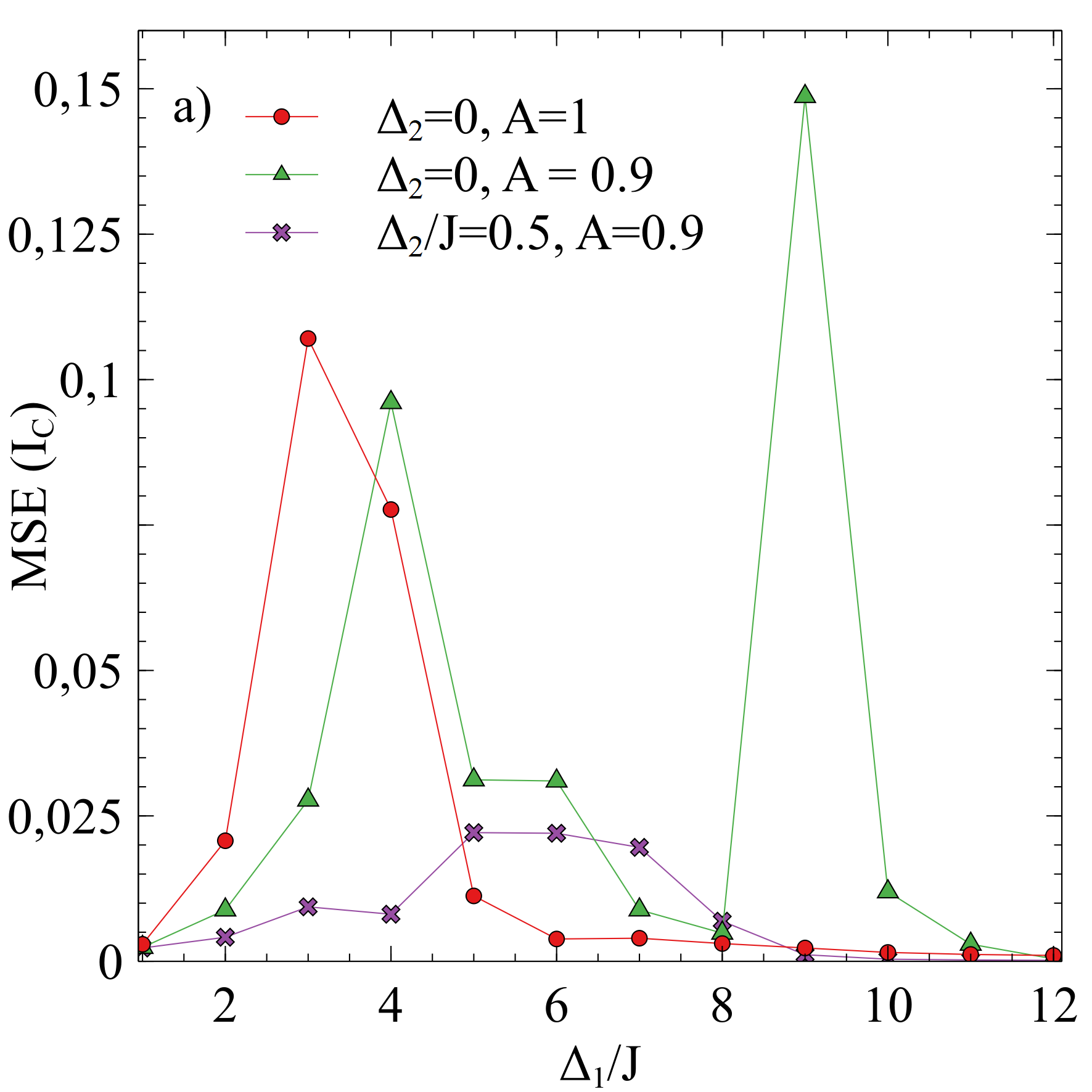}\includegraphics[scale=0.45]{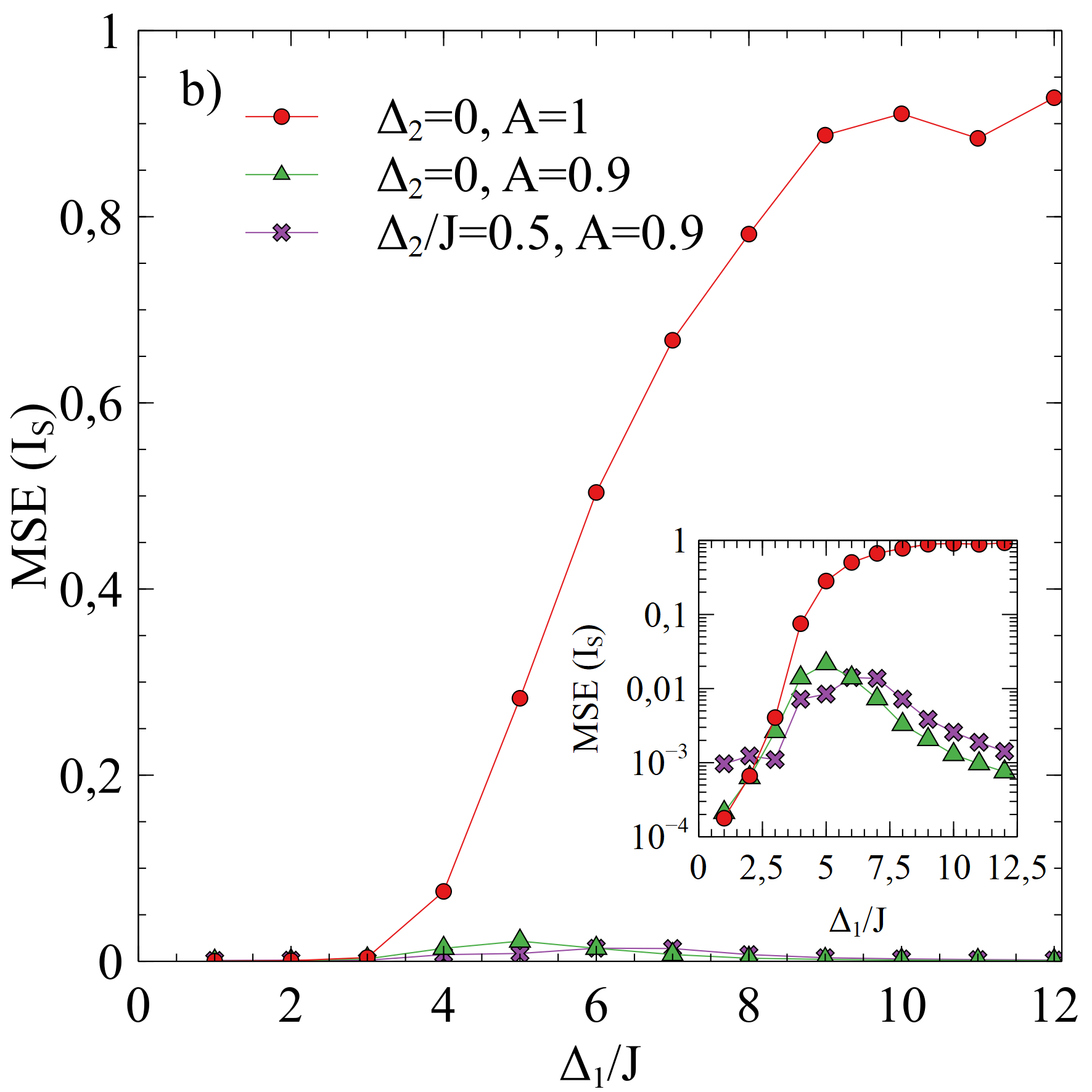}
\caption{Dependence of the mean square error MSE($I_{C/S}$) on the linear
potential strength $\Delta_{1}/J$ (for MSE($I_{C/S}$) definition,
see Eq. (\ref{eq: MSE(I)})). MSE($I_{C}$) and MSE($I_{S}$) are
calculated for charge (a) and spin (b) imbalance, respectively. Different
parameter ranges are considered, circles correspond to $A=1$,
$\Delta_{2}/J=0$, triangles to $A=0.9$, $\Delta_{2}/J=0$
and crosses to $A=0.9$, $\Delta_{2}/J=0.5$. In the inset of
(b) we plotted the same data as in (b) but with an additional logarithmic
scale in the vertical axis. The other parameters are the same as in
Fig. \ref{fig: imbalances time dependence}. \label{fig: MSE imbalance}}
\end{figure*}

\begin{figure*}
\includegraphics[scale=0.38]{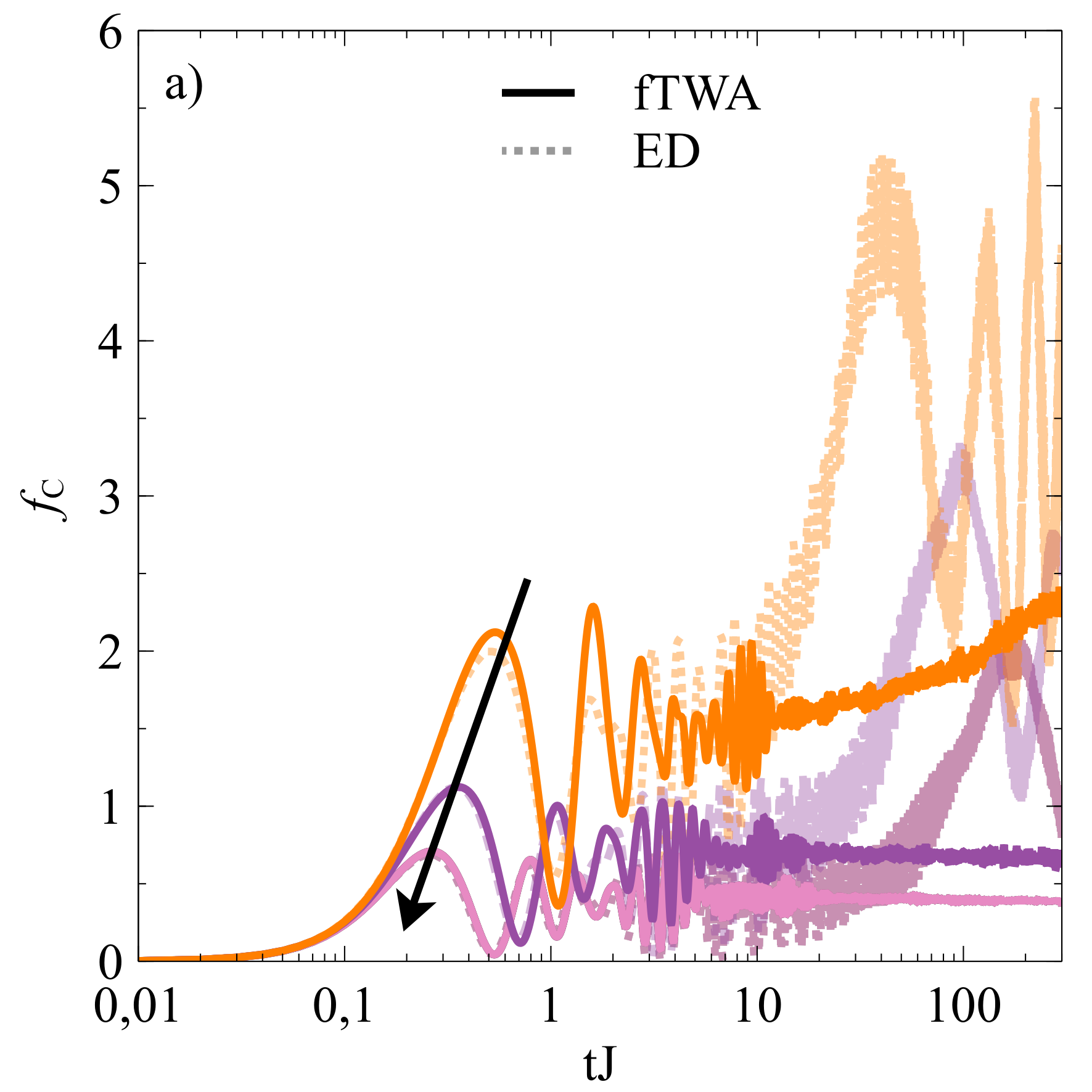}\includegraphics[scale=0.38]{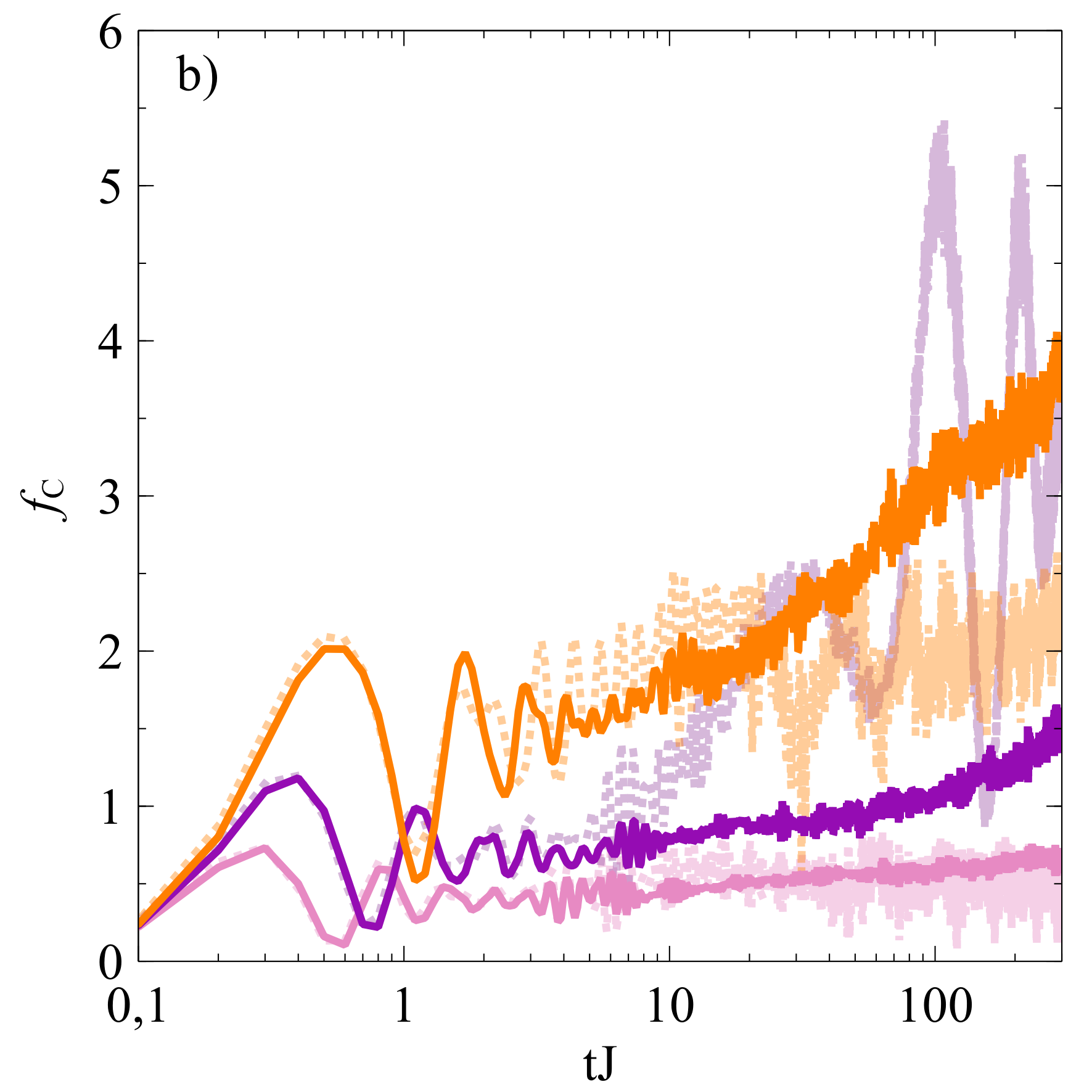}\includegraphics[scale=0.38]{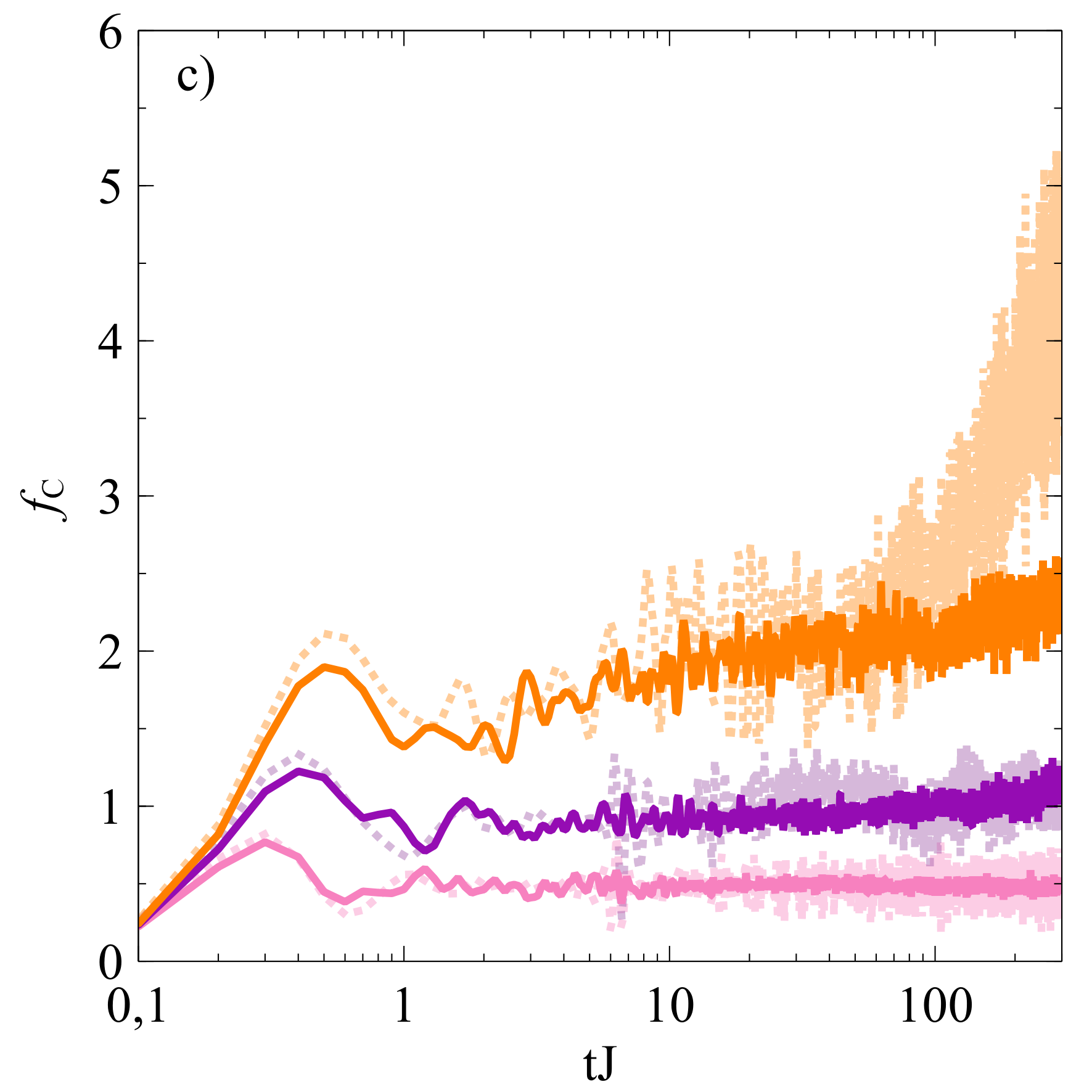}
\includegraphics[scale=0.38]{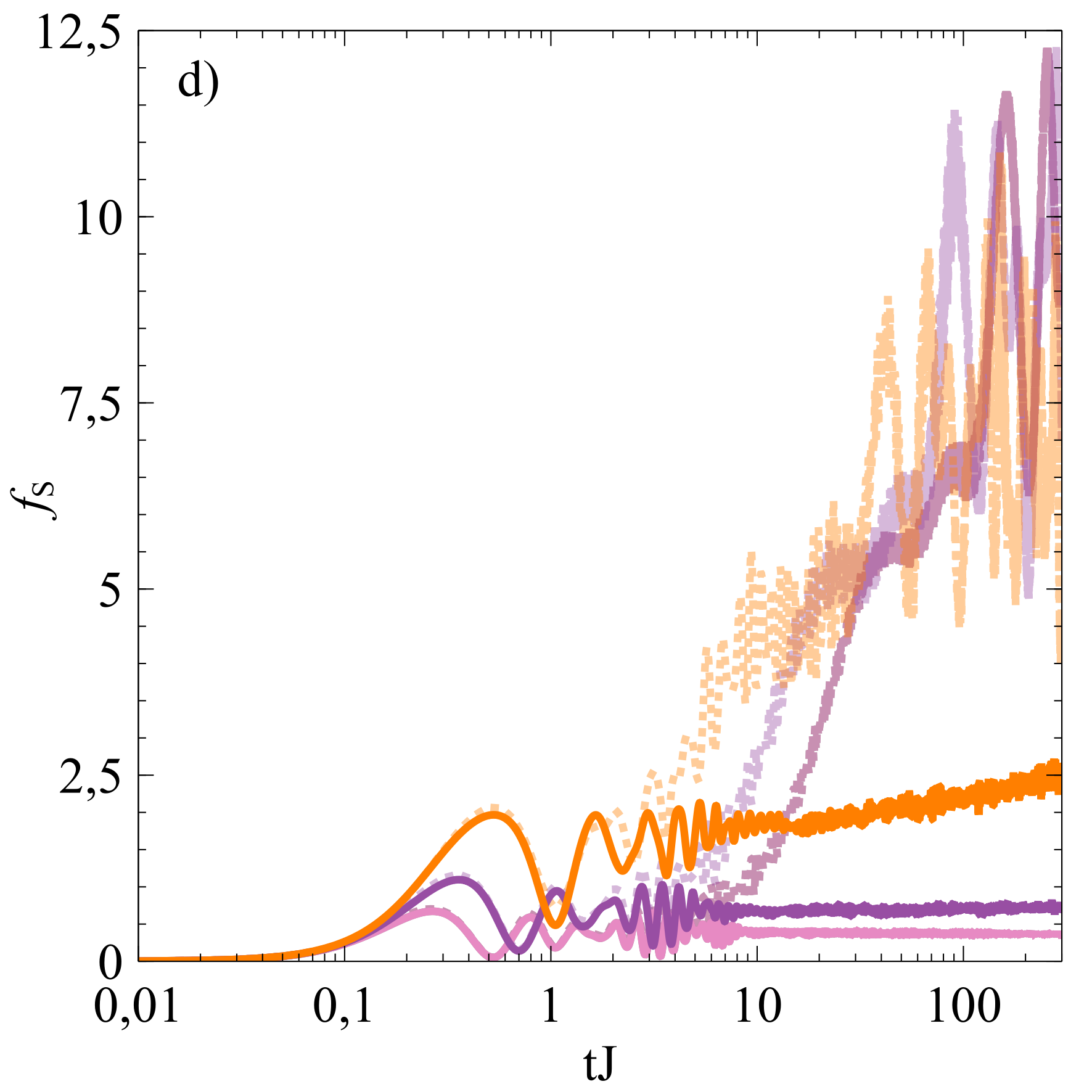}\includegraphics[scale=0.38]{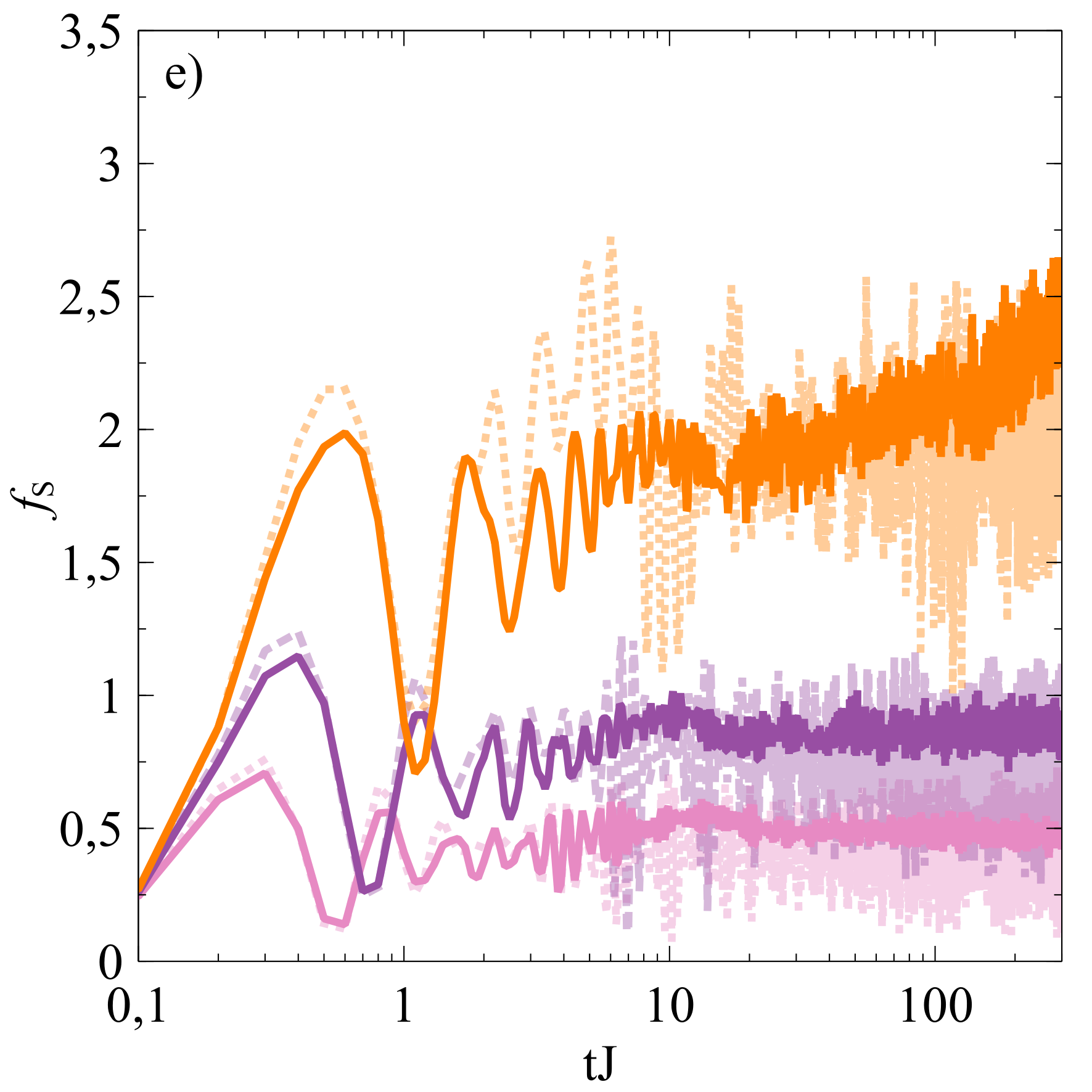}\includegraphics[scale=0.38]{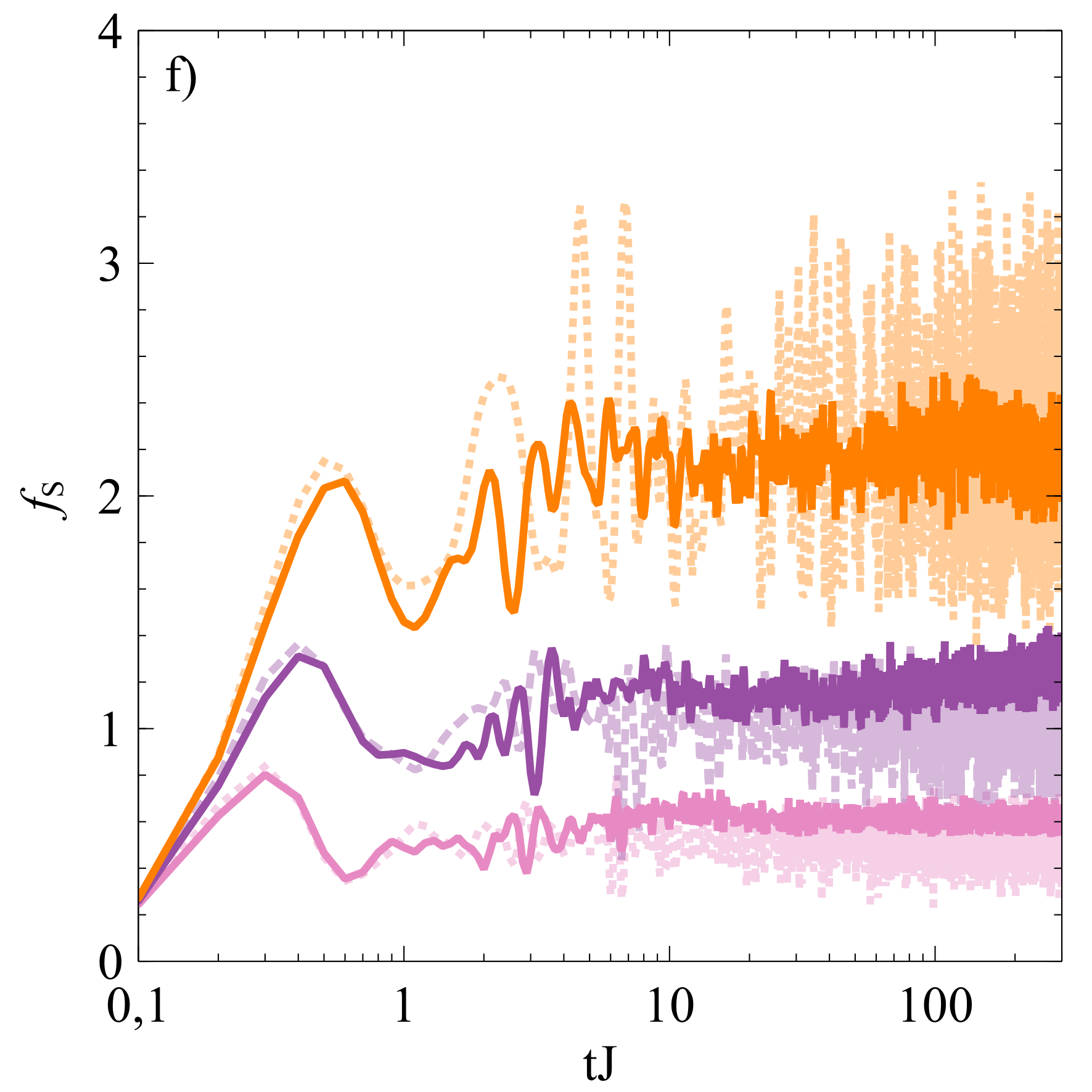}\caption{Time dependence of QFI for charges (Fig. a-c) and spins (Fig. d-f).
In each plot different strengths of the linear potential $\Delta_{1}/J$
are taken, i.e. $\Delta_{1}/J=6,\,9,\,12$ from the top to bottom (direction of increasing values of $\Delta_{1}/J$ is marked by the arrow in Fig. a). The other parameters are the same as in Fig.
\ref{fig: imbalances time dependence}. \label{fig: QFI time dependence}}
\end{figure*}

\begin{figure*}
\includegraphics[scale=0.45]{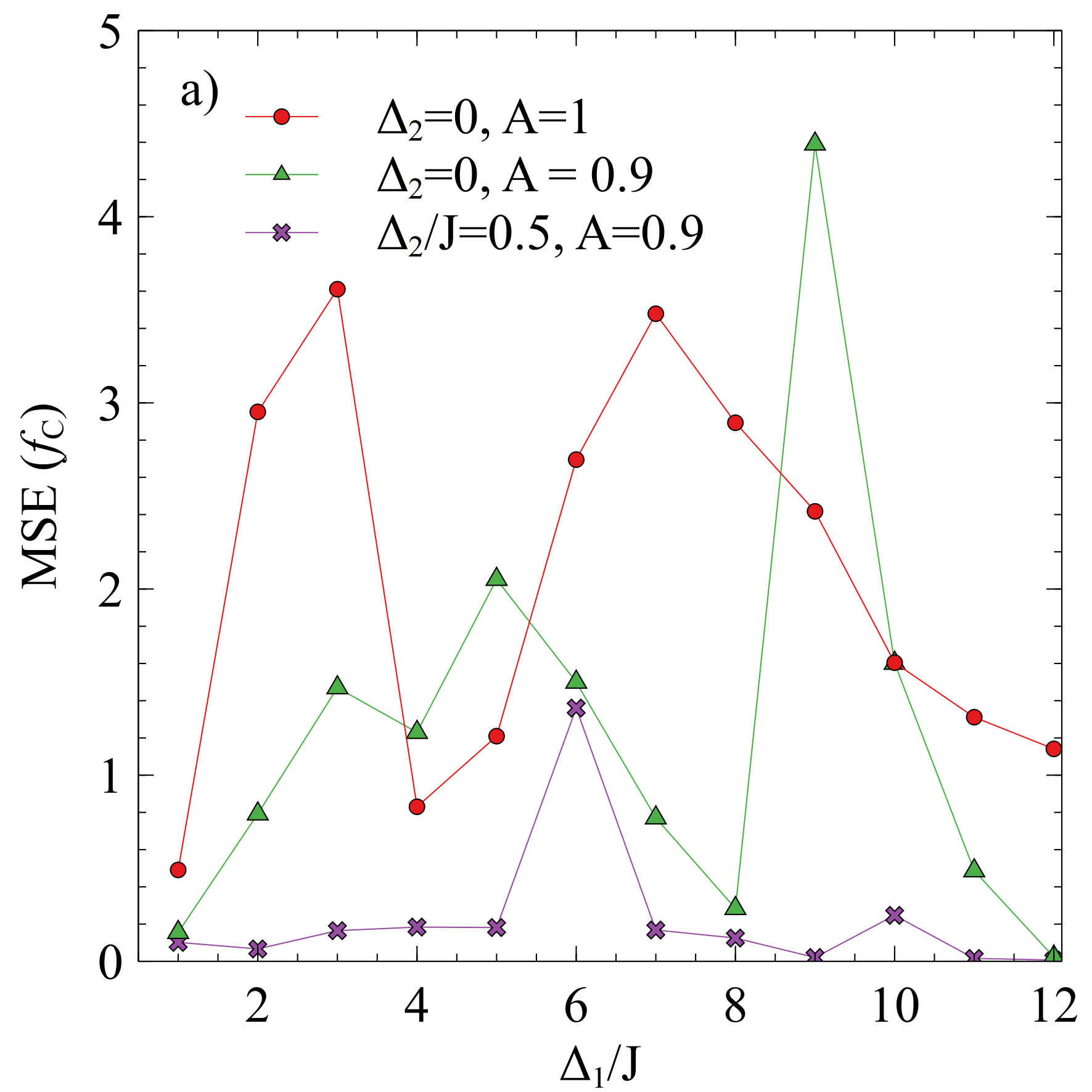}\includegraphics[scale=0.45]{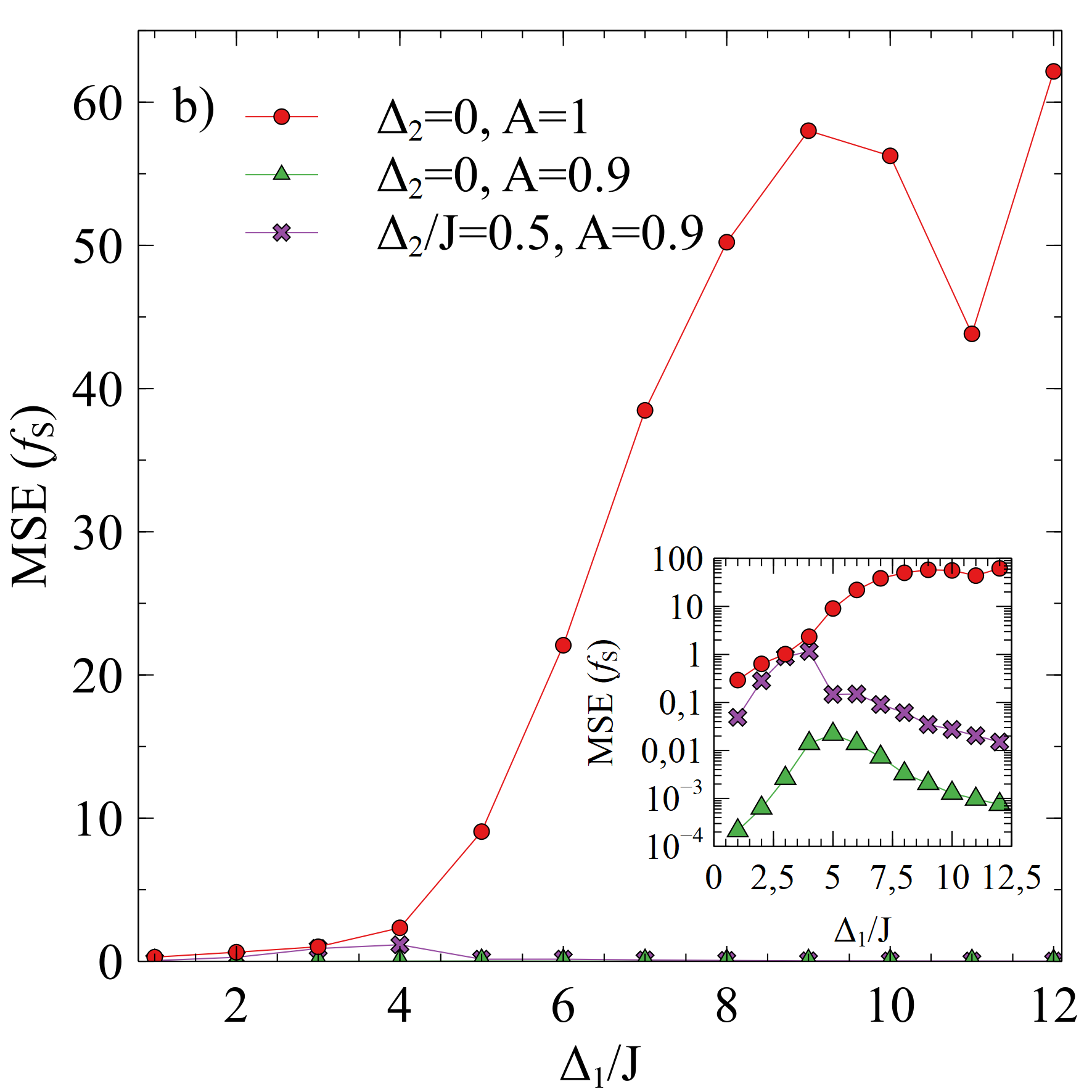}
\caption{Dependence of the MSE($f_{C/S}$) on the linear
potential strength $\Delta_{1}$. MSE($f_{C}$) and MSE($f_{S}$)
are calculated for charge (a) and spin (b) QFI, respectively. Different
parameter ranges are considered, circles correspond to $A=1$,
$\Delta_{2}/J=0$, triangles to $A=0.9$, $\Delta_{2}/J=0$
and crosses to $A=0.9$, $\Delta_{2}/J=0.5$. In the inset of
(b) we plotted the same data as in (b) but with an additional logarithmic
scale in the vertical axis. The other parameters are the same as in
Fig. \ref{fig: imbalances time dependence}.\label{fig: MSE QFI}}
\end{figure*}

To benchmark the fTWA method, we compare the results of semiclassical simulations with those of
ED in a finite one-dimensional system at half-filling (8 lattice sites are investigated).
The role of harmonic potential and spin dependence of the linear field is stressed by using
the imbalance functions and QFI. We chose these quantities because
they are accessible experimentally in trapped atoms and ions experiments
and are useful in a discussion of ergodicity breaking in different systems
\citep{mblschreiber,Choi2016,PhysRevLett.116.140401,PhysRevX.7.041047,Lschen2017,Smith2016,Smith2019}.

The imbalance function measures the distribution of charges (densities) and
spin degrees of freedom at a given time. Assuming that the system starts
from a charge density wave (CDW) where the even sites are doubly occupied
and the odd ones are empty, the imbalance $I_{C}$ is defined as
\begin{equation}
I_{C}=\frac{1}{N}\left(\left\langle \hat{C}_{e}\right\rangle -\left\langle \hat{C}_{o}\right\rangle \right),
\end{equation}
with
\begin{equation}
\hat{C}_{e/o}=\sum_{i\in\text{even/odd sites}}\hat{c}_{i},\label{eq: C}
\end{equation}
where $\hat{c}_{i}=\hat{n}_{i\uparrow}+\hat{n}_{i\downarrow}$ is the local charge density, $N$
is the number of fermions, $\hat{C}_{e}$ and $\hat{C}_{o}$ are the operators
of the total charge on even and odd sites, respectively.

Correspondingly, for the spin degrees of freedom, the imbalance function $I_{S}$
can be defined in the following way
\begin{equation}
I_{S}=\frac{1}{N}\left(\left\langle \hat{S}_{e}\right\rangle -\left\langle \hat{S}_{o}\right\rangle \right),
\end{equation}
with
\begin{equation}
\hat{S}_{e/o}=\sum_{i\in\text{even/odd sites}}\hat{s}_{i},\label{eq: S}
\end{equation}
where $\hat{s}_{i}=\hat{n}_{i\uparrow}-\hat{n}_{i\downarrow}$ is the local spin magnetization, $\hat{S}_{e}$
and $\hat{S}_{o}$ are the operators of the total spin magnetizaton (z component) on even and odd
sites, respectively. In order to study the dynamics of the spin degrees of
freedom we chose the initial spin density wave (SDW), i.e. even (odd)
sites containing fermions with spins up (down).

Moreover, to efficiently discuss a quantitative difference between
fTWA and ED, the mean square error (MSE) is analyzed, given
by the formula 
\begin{equation}
\text{MSE}(I_{C/S})=\frac{1}{N_{\text{s}}+1}\sum_{j=0}^{N_{\text{s}}}\left(I_{C/S}^{\text{ED}}(j\Delta t)-I_{C/S}^{\text{fTWA}}(j\Delta t)\right)^{2},\label{eq: MSE(I)}
\end{equation}
where $\Delta t = 0.01/J$ is the  time step after which data are numerically collected,
$N_{\text{s}}\Delta t=300/J$ is the total time of simulations, $C$ and $S$
indices correspond to the charge and spin channel, respectively. Correspondingly,
$I_{C/S}^{\text{ED}}$ and $I_{C/S}^{\text{fTWA}}$ stand for the imbalances
calculated by using the ED and fTWA methods.

In Fig. \ref{fig: imbalances time dependence} we plot the time dependences of the imbalances
$I_{C}$ and $I_{S}$ in the fTWA and ED simulations. We
first focus on the role of spin dependence of the linear potential. It is easily seen
that for a spin-independent potential, $A=1$ (see Fig. \ref{fig: imbalances time dependence}
a and d), delocalization of spin degrees of freedom takes place (Fig.
\ref{fig: imbalances time dependence} d). A similar behavior was previously observed in the context of the spin-independent disordered systems \citep{PhysRevB.94.241104,PhysRevLett.120.246602,PhysRevB.98.014203,PhysRevB.99.115111,PhysRevB.99.121110,PhysRevB.101.134203}.
In our simulations, this happens at times of the order of $\mathcal{O}(\text{tJ})$
and makes the fTWA to completely fail to describe the many-body quantum
dynamics in the intermediate and large linear potential strength limit
(see also the growth of MSE($I_{S}$) function in Fig. \ref{fig: MSE imbalance}
b). In Fig. \ref{fig: imbalances time dependence} e, we show that
introduction of a weak spin dependence of the linear potential, i.e. $A=0.9$, forbids
spin delocalization within the analyzed times and recovers the approximate
predictability of fTWA.

Having established an efficient description of the spin channel, we focus on
the role of harmonic potential in our semiclassical dynamics by setting
$\Delta_{2}/J=0.5$ (see, Fig. \ref{fig: imbalances time dependence}
c and f). Then the situation is reversed to that of the spin channel.
We observe enhancement of fTWA prediction in the charge channel which
is explicitly seen in MSE($I_{C}$) for intermediate and large
linear potential strength (see, Fig. \ref{fig: MSE imbalance} a).

In our studies we also look at the QFI which is a higher order correlation function in comparison
to imbalance (QFI is proportional to the variance of $\hat{C}_{e}-\hat{C}_{o}$
or $\hat{S}_{e}-\hat{S}_{o}$). For pure initial states analyzed
here, i.e. for CDW and SDW, the corresponding normalized QFI for charges
$f_{C}$ and spins $f_{S}$ has the form \citep{PhysRevLett.72.3439,PhysRevA.85.022321,PhysRevA.85.022322,Hauke2016}
\begin{equation}
f_{C}=\frac{4}{N}\left[\left\langle \left(\hat{C}_{e}-\hat{C}_{o}\right)^{2}\right\rangle -\left\langle \hat{C}_{e}-\hat{C}_{o}\right\rangle ^{2}\right],
\end{equation}
\begin{equation}
f_{S}=\frac{4}{N}\left[\left\langle \left(\hat{S}_{e}-\hat{S}_{o}\right)^{2}\right\rangle -\left\langle \hat{S}_{e}-\hat{S}_{o}\right\rangle ^{2}\right].
\end{equation}
Similarly as in the imbalance case we focus on the three regimes:
(\textit{i}) with spin-independent tilt ($A=1$) and without a harmonic
potential ($\Delta_{2}=0$), see Fig. \ref{fig: QFI time dependence}
a and d, (\textit{ii}) with spin-dependent tilt ($A=0.9$) and without a
harmonic potential ($\Delta_{2}=0$), see Fig. \ref{fig: QFI time dependence}
b and e, (\textit{iii}) with spin-dependent tilt ($A=0.9$) and with a
harmonic potential ($\Delta_{2}=0.5$), see Fig. \ref{fig: QFI time dependence}
c and f. The predictability of fTWA for QFI in (\textit{i})
case is even worse than for the imbalance function. The abrupt increase in
QFI at later times is not properly described in terms of semiclassical description.
However introduction of the spin-dependent tilt and a harmonic potential
substantially improves the fTWA method. This conclusion is better illustrated
in Fig. \ref{fig: MSE QFI} where MSE($f_{C/S}$) is plotted (definition
of MSE($f_{C/S}$) corresponds to that given in Eq. (\ref{eq: MSE(I)})
for imbalance). We observe that $f_{C}$ is mostly improved for the
case $A=0.9$ and $\Delta_{2}=0.5$, while for $f_{S}$ the highest
enhancement of the fTWA method is observed in the case of the spin-dependent linear potential. The latter behavior is consistent with that
of the imbalance function for a spin channel (cf Fig. \ref{fig: imbalances time dependence}
d, f).

Interestingly, in the systems with a spin-dependent linear field and with an additional
harmonic potential (($iii$) regime), the MSE of imbalance functions and QFI show a peak at the
intermediate value of the linear potential strength. It means that fTWA gives the best prediction of quantum dynamics for weak and strong tilts. Such a feature was previously also observed for disordered systems when the disorder strength was varied \citep{PhysRevA.96.033604,PhysRevA.102.033338}. Moreover, we also noticed that in the ($iii$) regime and charge channel, fTWA imbalances decay faster than the corresponding ones in ED, which suggests that fTWA dynamics can be regarded as an upper bound for relaxation rates. This situation is similar to that of disordered systems studied recently for spinless interacting fermions~\cite{Iwanek2022}.

\section{Semiclassical dynamics of a two-dimensional system \label{sec: 2D}}

\begin{figure*}[t]
\includegraphics[scale=0.4]{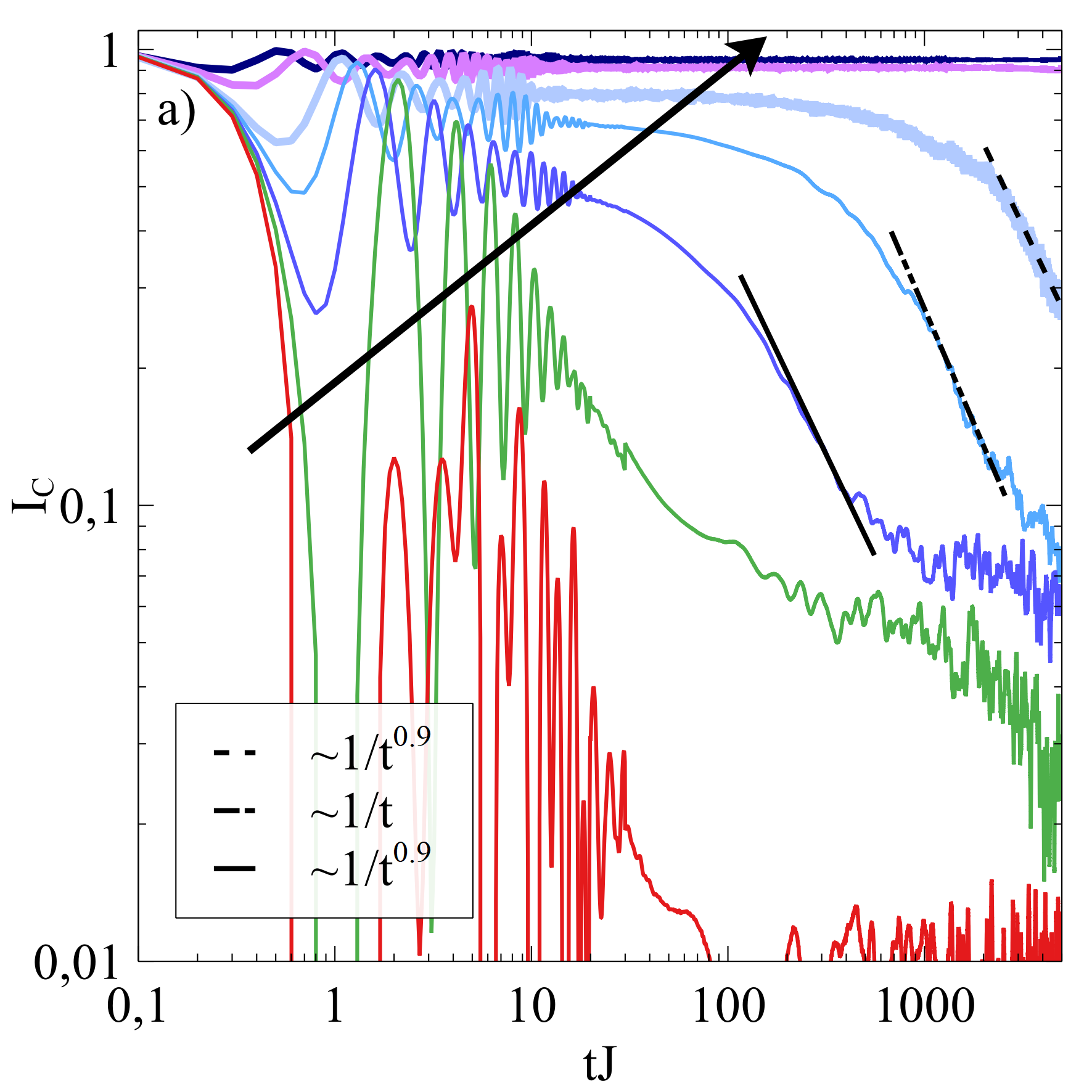}\includegraphics[scale=0.4]{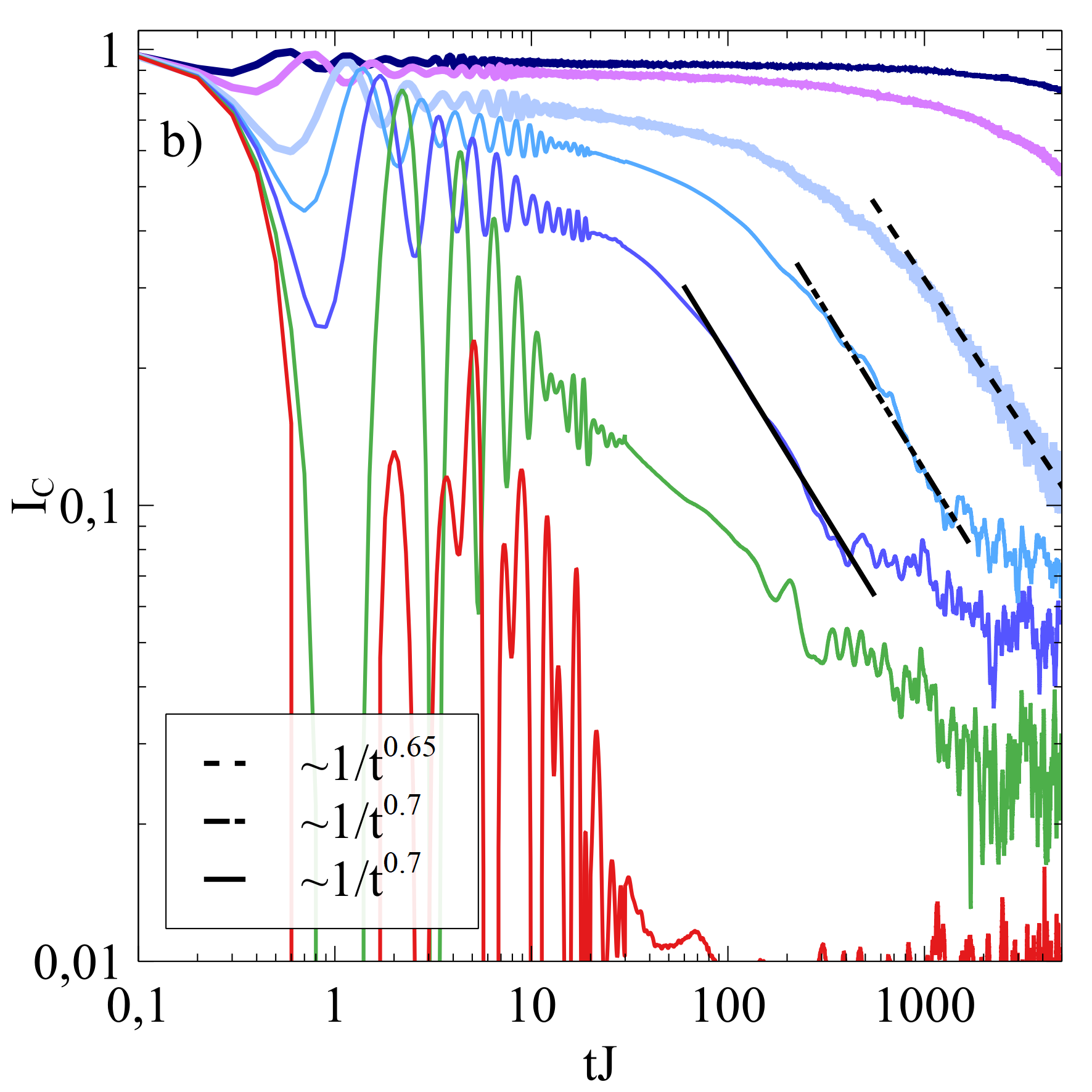}\includegraphics[scale=0.4]{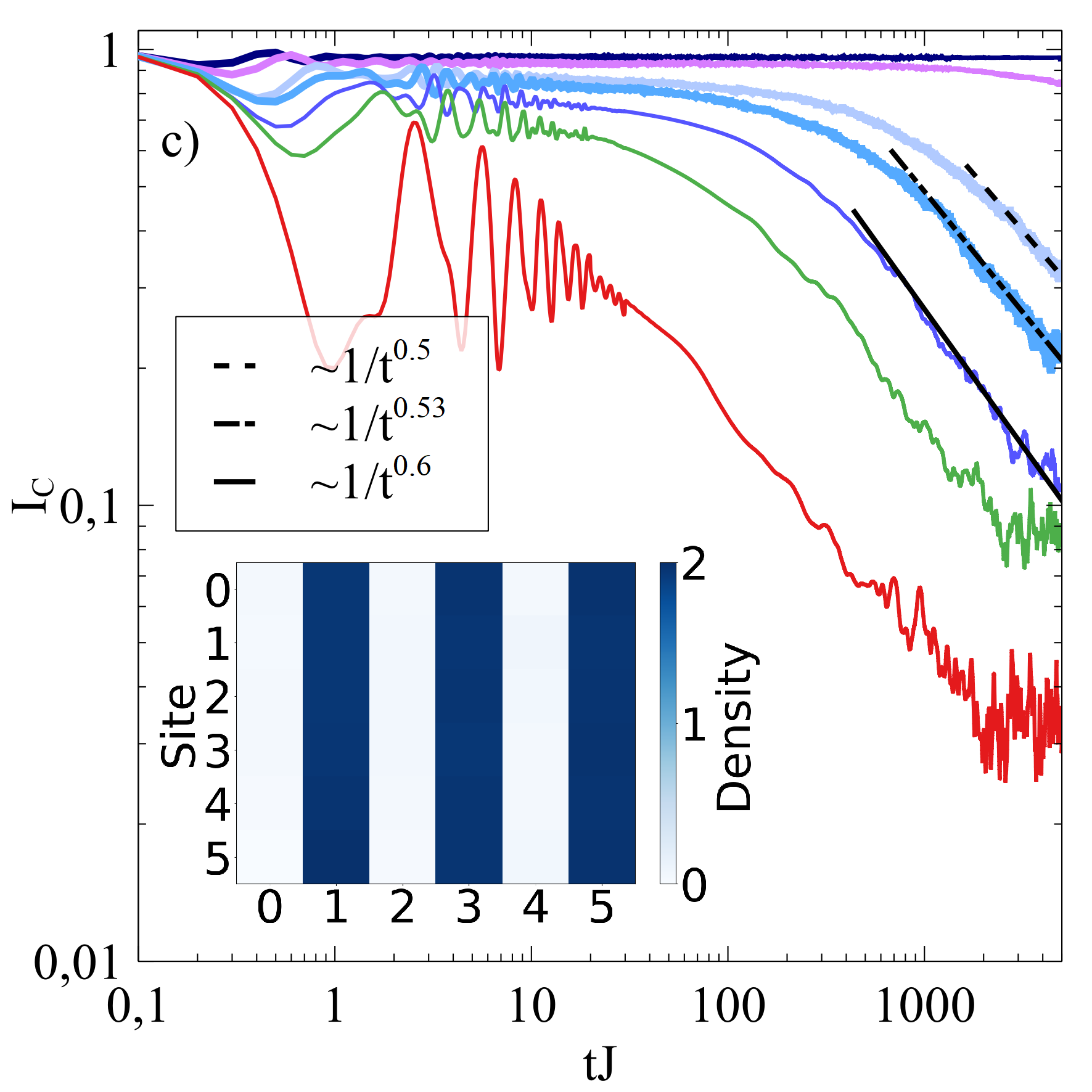}
\includegraphics[scale=0.4]{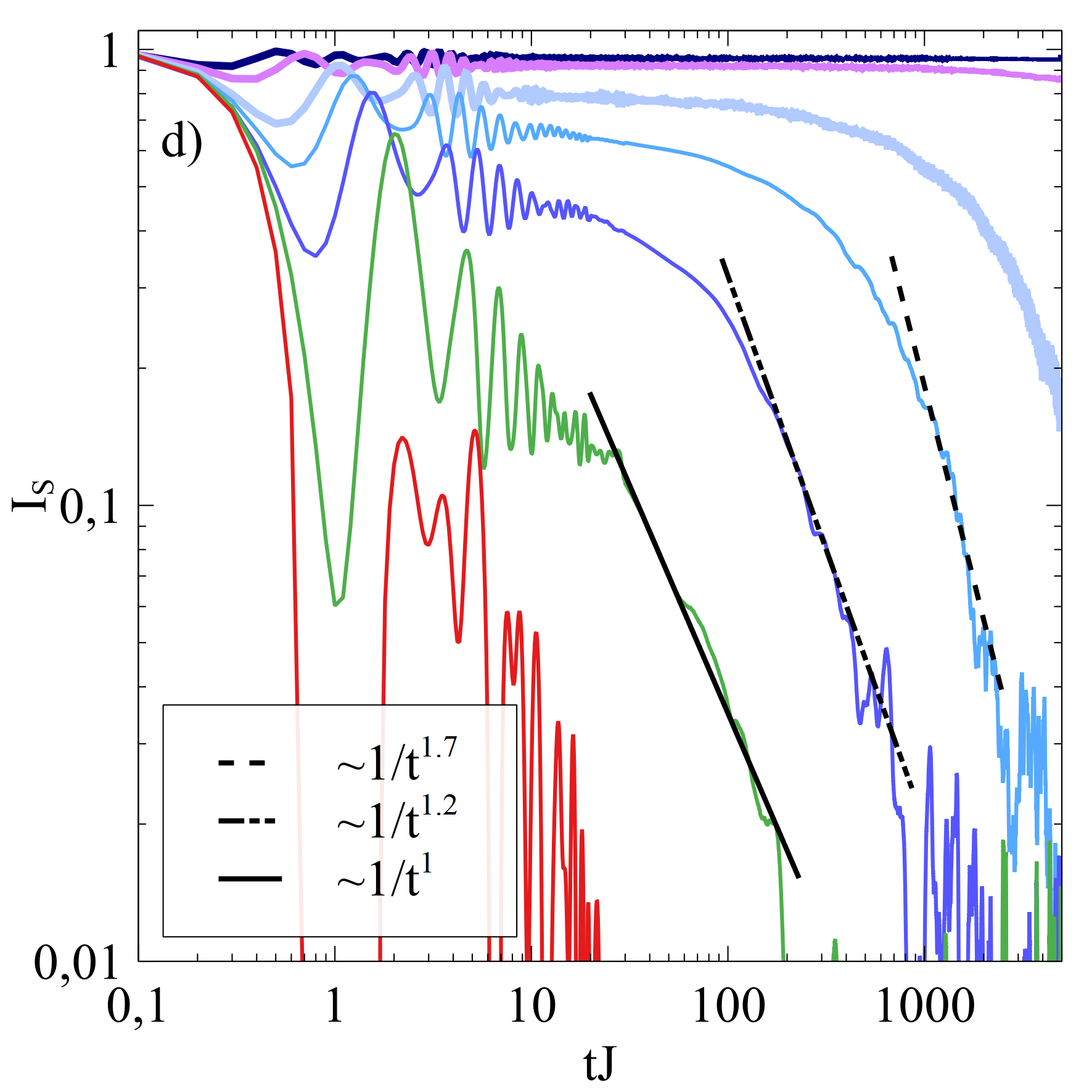}\includegraphics[scale=0.4]{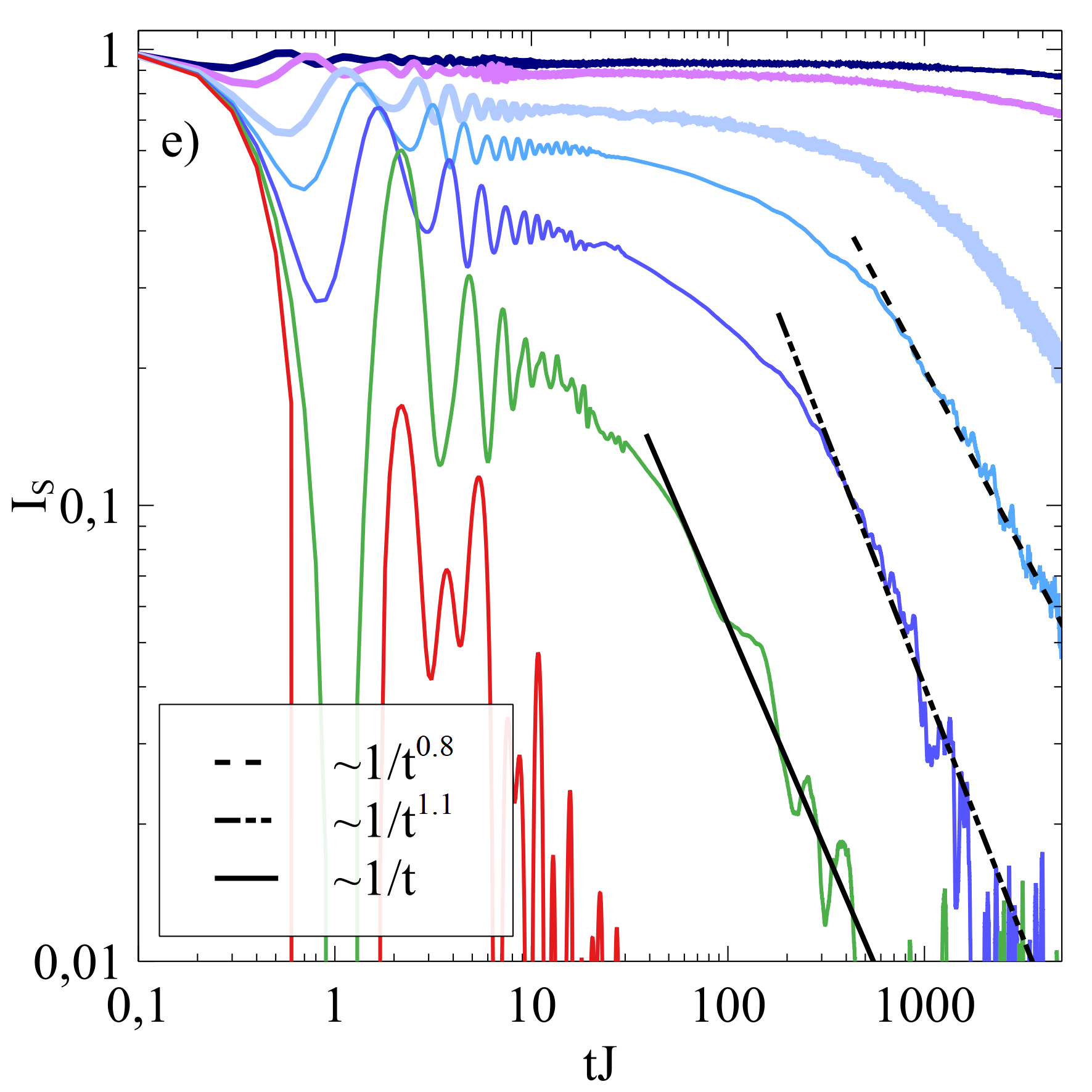}\includegraphics[scale=0.4]{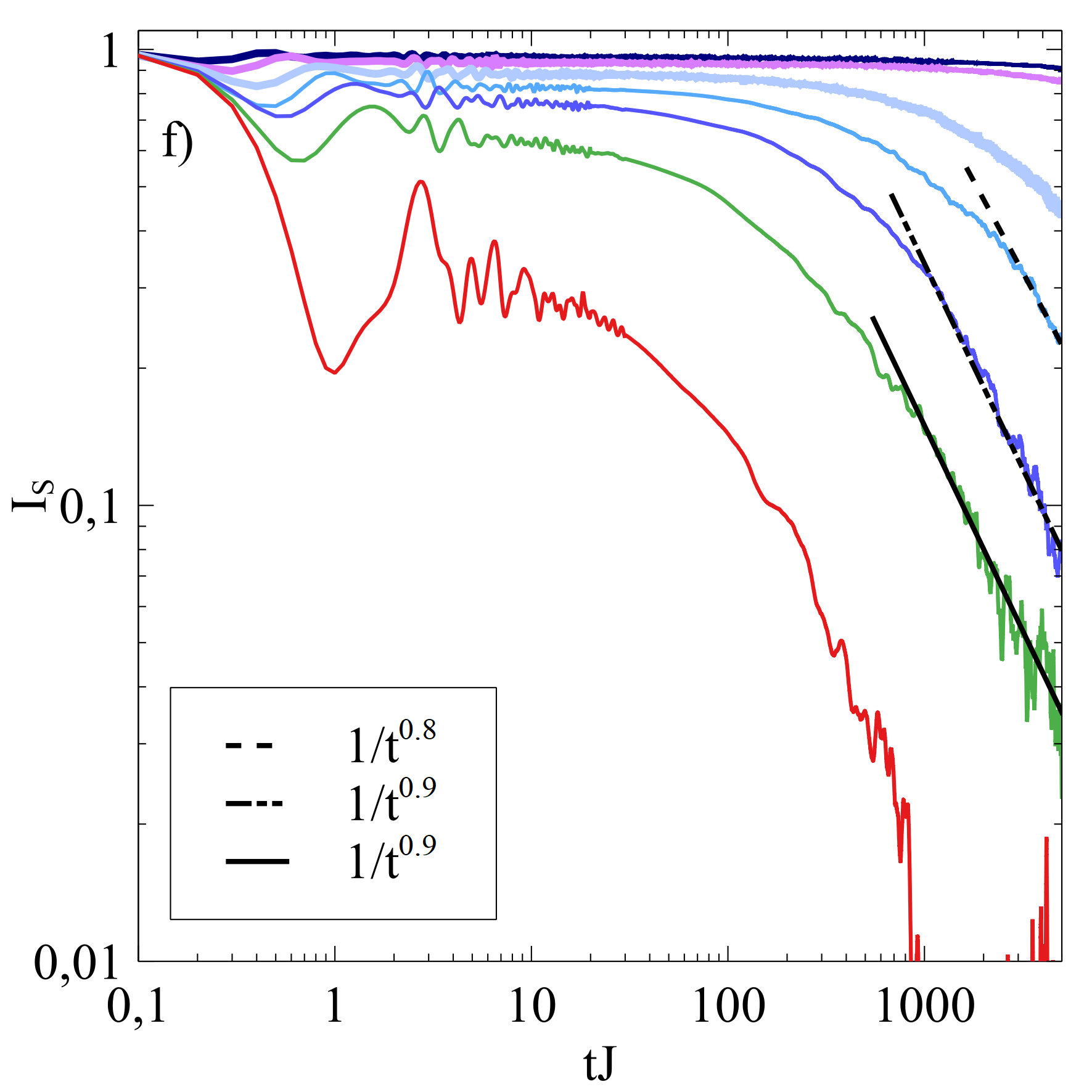}\caption{Time dependences of charge (Fig. a-c) and spin (Fig. d-f) imbalance
functions. In each plot, different strengths of the linear potential $\Delta_{1}/J$
are taken, i.e. $\Delta_{1}/J=1,\,3,\,4,\,5,\,6,\,9,\,12$ from the bottom
to top (direction of increasing values of $\Delta_{1}/J$ is marked by the arrow in Fig. a). The first column (a and d) corresponds
to $A=1$, $\Delta_{2}=0$, the second column (b and e) to $A=0.9$, $\Delta_{2}=0$,
the third column (c and f) to $A=0.9$, $\Delta_{2}/J=0.5$. Simulations
are performed for a finite two-dimensional system with $6\times6$ sites
and with the striped CDW (a-c) or SDW (d-f) initial conditions.
The structure of stripe-like initial conditions is presented in the
inset of Fig. (c). The other parameters are $U/J=1$, $j_{x,0}=$ $j_{y,0}=0$,
the number of trajectories used in fTWA is around 100. \label{fig: imbalance 2d}}
\end{figure*}

\begin{figure*}[t]
\includegraphics[scale=0.5]{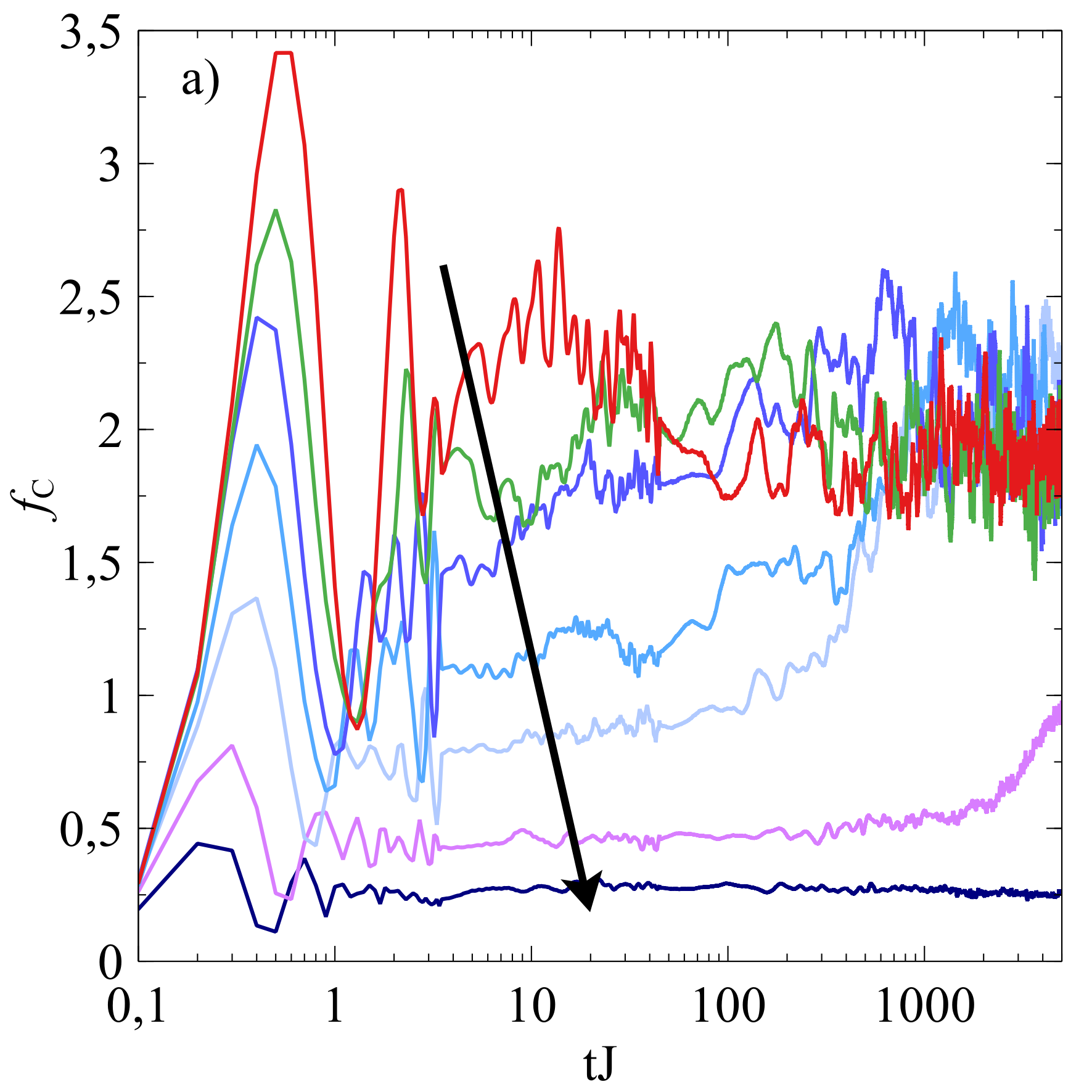}\includegraphics[scale=0.5]{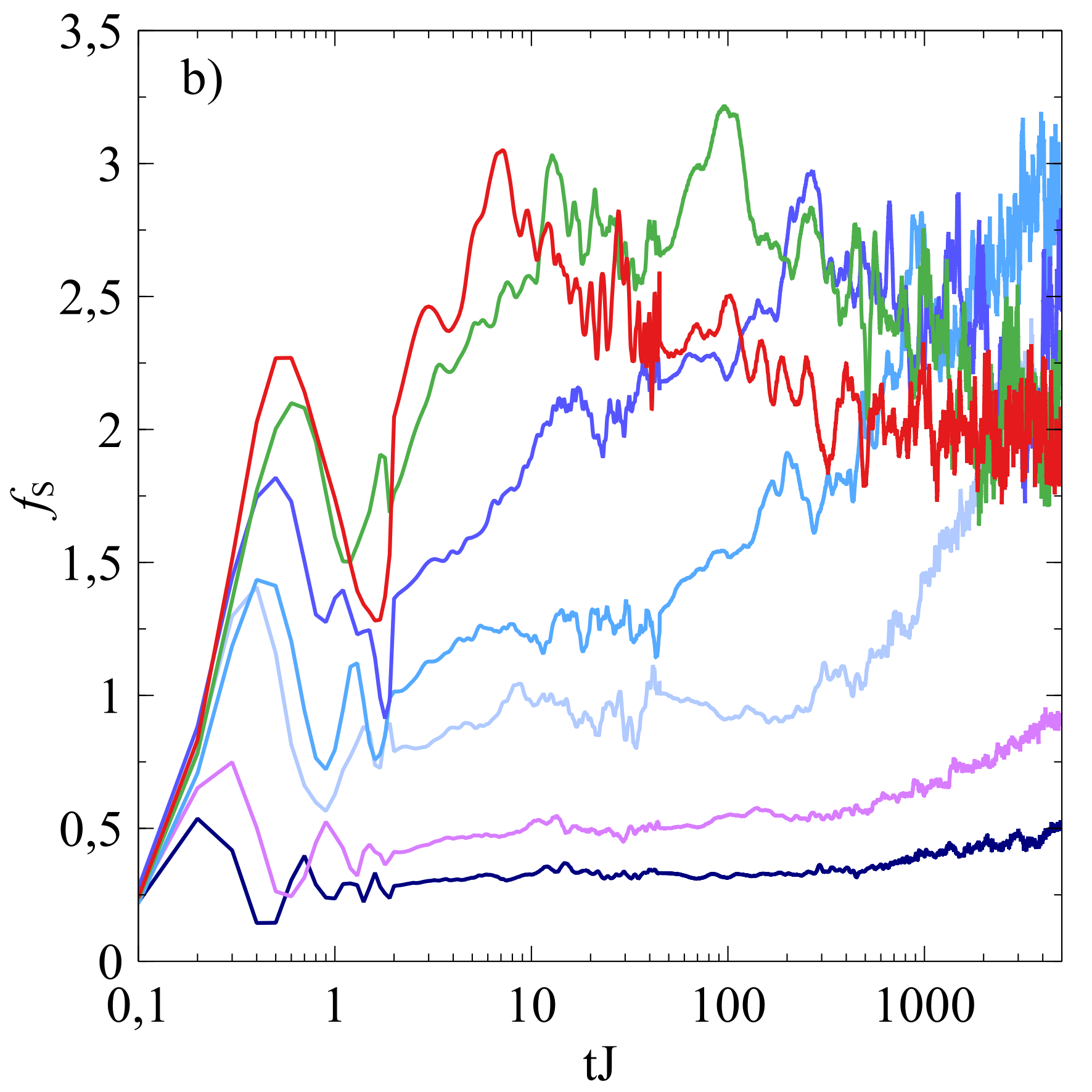}
\caption{Time dependence of QFI for charges (a) and spins (b). The other parameters
and denotation are the same as in Fig. \ref{fig: imbalance 2d}, however
here, increasing values of $\Delta_{1}/J$ correspond to the curves from the
top to bottom (direction of increasing values of $\Delta_{1}/J$ is marked by the arrow in Fig. a).\textcolor{red}{{} \label{fig: QFI 2d}}}
\end{figure*}

\begin{figure*}[t]
\includegraphics[scale=0.3]{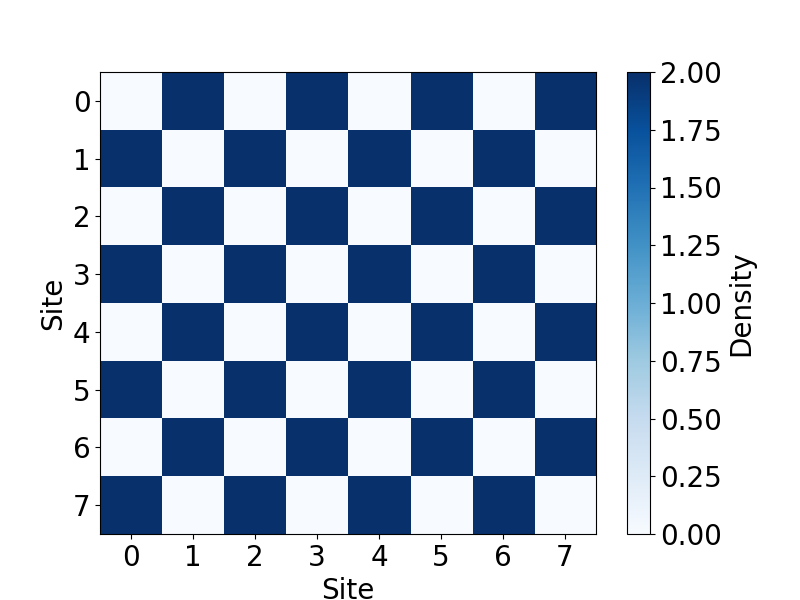}\includegraphics[scale=0.3]{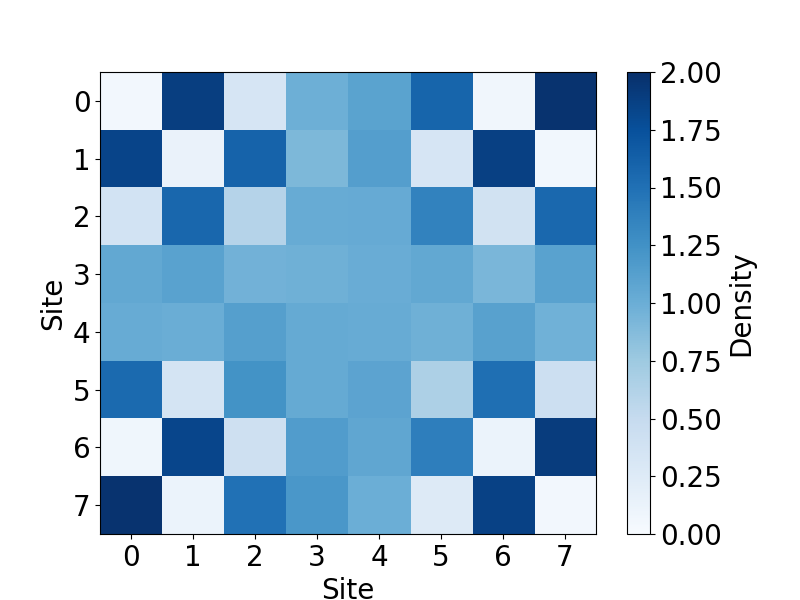}\includegraphics[scale=0.3]{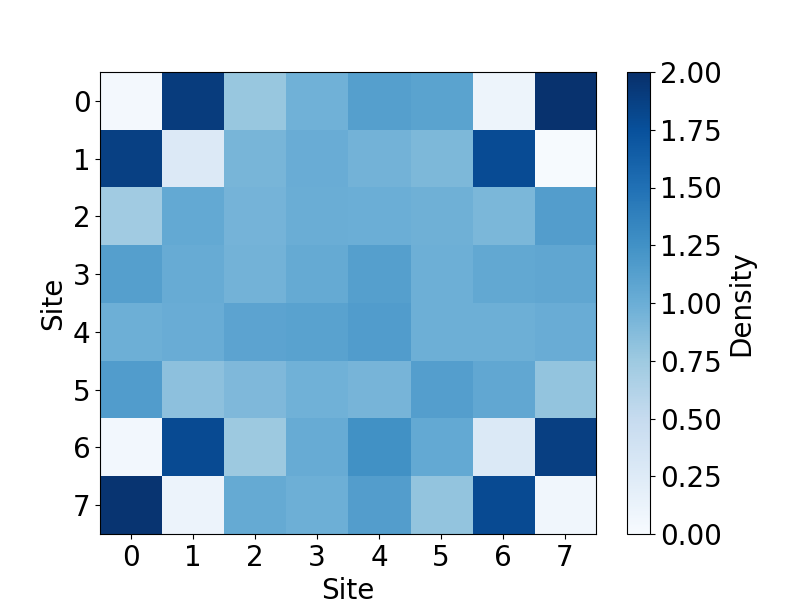}
\caption{Density plots of charge distribution for $8\times8$ square lattice
at different times $tJ=0$ (left), $tJ=10$ (center), $tJ=300$ (right).
In simulations $\Delta_{1x}=\Delta_{1y}=\Delta_{1}$ is chosen and around 100 fTWA trajectories are used in each plot. The other parameters are $U/J=1$, $A=0.9$, $\Delta_{1}/J=-6$, $\Delta_{2}/J=2$,
$j_{0x}=j_{0y}=2$.\textcolor{red}{{} \label{fig: checkerboard}}}
\end{figure*}

After benchmarking fTWA against the exact simulations, we focus our analysis
on the system sizes that are beyond the reach of ED. Using the advantage offered by the fTWA method, that is the fact that it can be easily extended to higher dimensional systems, we focus on the system with a square lattice.
It is worth mentioning that higher dimensional lattices have a higher
coordination number and we expected that performance of fTWA, as a semiclassical method, would be improved.

In Sec. \ref{sec: benchmark} it was shown that the predictability of
fTWA for the case with spin-independent linear potential can be questioned
for longer times (especially as far as the spin degrees of freedom are concerned). However,
we decided to include this case in our two-dimensional simulations
in order to show that the slowing down of dynamical behavior through
the spin dependence of the linear potential is also clearly observed in the semiclassical
picture in two dimensions (similarly like in the ED case analyzed for
Hubbard chain in the previous section).

For a two-dimensional lattice system, the on-site potential $\Delta(j,\sigma)$ which was
introduced in Sec. \ref{sec: fTWA-for-Hubbard}, has to be generalized
to the two spatial directions, i.e.
\begin{align}
\Delta(j,\sigma) & =\Delta_{1x}\left(\delta_{\sigma\downarrow}+A\delta_{\sigma\uparrow}\right)j_{x}+\Delta_{2x}(j_{x}-j_{x,0})^{2}\nonumber \\
 & +\Delta_{1y}\left(\delta_{\sigma\downarrow}+A\delta_{\sigma\uparrow}\right)j_{y}+\Delta_{2y}(j_{y}-j_{y,0})^{2}\label{eq: on-site potential in two-dimension}
\end{align}
where $j=(j_{x},j_{y})$ is now a vector indicating the location of a given
lattice site (the $j_{x}$ and $j_{y}$ are the Cartesian coordinates in the $x$
and $y$ directions, respectively). The coordinates $j_{x,0}$ and $j_{y,0}$ denote
the center of the harmonic potential. In order to avoid lattice directions
for which there is no potential change, throughout most of the work,
we assume that the strengths of linear potentials in the $x$ and $y$
directions are $\Delta_{1x}=\Delta_{1}$ and $\Delta_{1y}=\sqrt{2}\Delta_{1}$,
respectively \citep{vanNieuwenburg2019}. However, for simplicity,
the harmonic potential strength satisfies the condition $\Delta_{2x}=\Delta_{2y}=\Delta_{2}$.
In Eq. (\ref{eq: on-site potential in two-dimension}) we also assumed
that the spin dependence of the linear potential given by the parameter $A$ is the same for $x$ and
$y$ lattice dimensions.

Firstly we analyze the behavior of imbalances $I_{C}$ and $I_{S}$ at
long times for $6\times6$ lattice. We set the initial conditions in the form
of stripes (see inset in Fig. \ref{fig: imbalance 2d} c) which are
directly accessible in ultracold atom experiments \citep{PhysRevLett.116.140401,PhysRevX.7.041047}.
In the striped CDW initial state every second stripe is doubly occupied
and the others are empty (in striped SDW every second stripe contains fermions
with spin up and the other sites are filled with fermions with spin down).
The choice of such initial conditions needs a comment because the definitions
of $I_{C}$ and $I_{S}$ given in Sec. \ref{sec: benchmark} have
to be updated. Instead of Eq. (\ref{eq: C}) and (\ref{eq: S}), we introduce
the following definitions of $\hat{C}$ and $\hat{S}$ operators
\begin{equation}
\hat{C}_{e/o}=\sum_{i\in\text{X}_{e/o}}\hat{c}_{i},\ \hat{S}_{e/o}=\sum_{i\in\text{Y}_{e/o}}\hat{s}_{i},
\end{equation}
where $\text{X}_{e}$ ($\text{Y}_{e}$) and $\text{X}_{o}$ ($\text{Y}_{o}$)
denote the sets of sites that are initially doubly occupied (fermions
with spin up) and empty (fermions with spin down), respectively.

The outcome of the numerical simulations of $I_{C}$ and $I_{S}$ are presented in Fig.
\ref{fig: imbalance 2d} in which the results of three physical situations corresponding
to those in Fig. \ref{fig: imbalances time dependence} are plotted.
In the simulations parameters are chosen in such a way that the imbalance function without a tilt potential decays
near zero suggesting ergodic behavior within the analyzed time scales.
In each of the three cases (\textit{i-iii}, see Sec. \ref{sec: benchmark}),
as expected, imbalance dynamics for $I_{C}$ and $I_{S}$ are slowing
down when the strength of tilt is increased. We also observe that the relaxation
of imbalances, after introducing a harmonic and spin-dependent linear potential,
becomes slower for weak and intermediate tilts. To be more
specific, in the charge channel and with the spin-independent linear potential ($A=1$), nearly diffusive
dynamics of densities is observed at longer times, i.e. $I_{C}\sim t^{-\gamma}$
where $\gamma=1$ (Fig. \ref{fig: imbalance 2d} a). After introducing
the spin dependence of the linear field ($A=0.9$), the subdiffusive behavior appears ($\gamma<1$) which is
further strengthened by a harmonic potential (Fig. \ref{fig: imbalance 2d}
b and c). It is worth noting that the subdiffusive behavior was also observed for two-dimensional
interacting systems with a sufficiently strong disorder \citep{PhysRevA.102.033338,BarLev2016}.
For the spin degrees of freedom, the situation is more complex due to spin
dependence of the linear potential. For the spin-independent tilt, spin transport is superdiffusive
and approach diffusive when the spin dependence is imposed (\ref{fig: imbalance 2d}
d and e). Introduction of a harmonic potential makes the spin dynamics become subdiffusive,
similarly like in the charge case. This is especially visible for higher values of the linear potential
strength, see Fig. \ref{fig: imbalance 2d} c and f. Interestingly, the subdiffusive behavior of spin degrees of freedom was also observed in the disordered two-dimensional Hubbard model \citep{PhysRevA.102.033338}.
It is also worth mentioning, that some delocalization features of
the initial striped CDW state with short-wavelength have been also recently
reported in Ref. \citep{PhysRevB.105.134204}. This is consistent
with our studies, however, in Ref. \citep{PhysRevB.105.134204}, different
tilt direction and shorter time scales have been analyzed, therefore,
direct comparison needs further investigation which we left for future
studies.

It is important to mention that the fitting curves $t^{-\gamma}$ in Fig,
\ref{fig: imbalance 2d} were obtained for the long-time limit and for three
fixed values of $\Delta_{1}/J$. It is straightforward to notice
that in Fig. \ref{fig: imbalance 2d} a and b there are significant
deviations from these fitting curves at later times. It can be accounted for by
significant finite-size effects in the dynamics, which is faster in
the system without a harmonic potential, see Appendix \ref{sec:Appendix:-Finite-size-effects}.
In Appendix \ref{sec:Appendix:-Finite-size-effects} we also explain
that the finite size effects can be neglected in the cases when fTWA gives
the lowest errors and when it mimics the behavior of disordered systems.

We also investigate $f_{C}$ and $f_{S}$ focusing on the limit
in which fTWA satisfactorily describes the long-time dynamics, i.e. when
the spin dependence ($A=0.9)$ and the harmonic potential ($\Delta_{2}=0.5$)
are introduced. The results are presented in Fig. \ref{fig: QFI 2d} for
charge and a spin channels. Interestingly, in both situations we observe
a logarithmic-like growth of QFI, which is slower for higher values of the linear
potential. Here again as for imbalances, the dynamics of QFI is similar to that of
strongly disordered systems in one and two dimensions \citep{PhysRevB.99.241114,Guo2020,PhysRevA.102.033338}
or that of tilted triangular ladders \citep{PhysRevLett.127.240502}. In Appendix
\ref{sec:Appendix:-Finite-size-effects} we also show that the finite-size
effects do not play a significant role in the logarithmic growth and can
be neglected.

In the end of this section we also look at the competition between the
linear ($\Delta_{1}$) and harmonic ($\Delta_{2}$) potentials in the parameter
range in which additional harmonicity of the lattice leads to the appearance
of long-lived ergodic and non-ergodic regions. Fig. \ref{fig: checkerboard}
presents the density plot of charge distribution on the lattice. At the expenses of shorter time analysis, we increase the size of the lattice
to $8\times8$ and set the lowest value of harmonic potential at the lattice
center. To precisely catch the density decay on the individual sites we
choose a checkerboard-like structure of the initial CDW-like state in which
only charge channel is analyzed. In the presented simulations the linear potential is three times stronger than the harmonic and we simply
choose $\Delta_{1x}=\Delta_{1y}=\Delta_{1}$. We observed that within
the analyzed time scales, charges represented by doubly occupied sites,
did not decay at the corners of the lattice. The corresponding phase separation
has been also recently observed in the one-dimensional system in which the
effective local field was used for explanation of such behavior \citep{Yao2021,PhysRevResearch.2.032039,PhysRevB.102.104203,Morong2021-mh}.

\section{Summary and outlook \label{sec: Summary}}

\begin{figure*}
\includegraphics[scale=0.3]{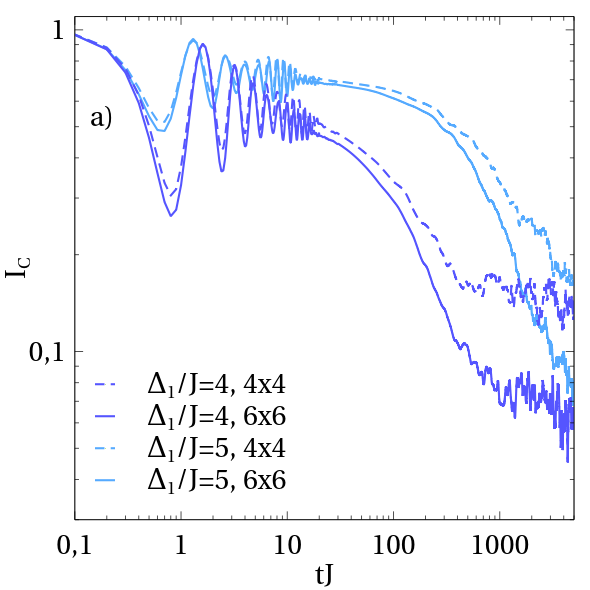}\includegraphics[scale=0.3]{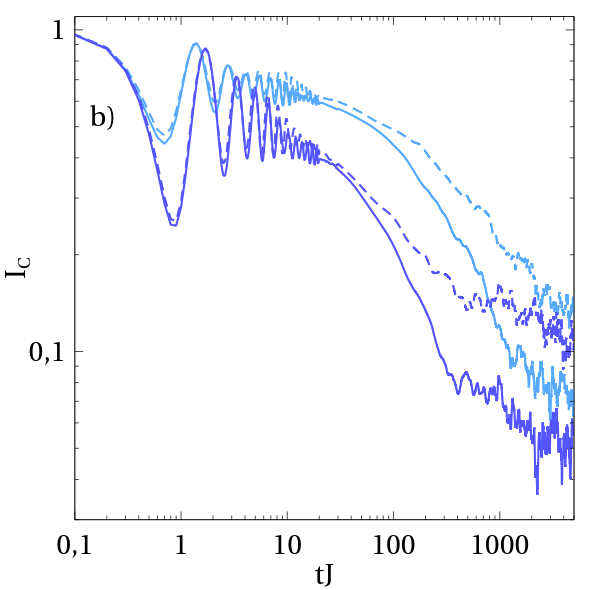}\includegraphics[scale=0.3]{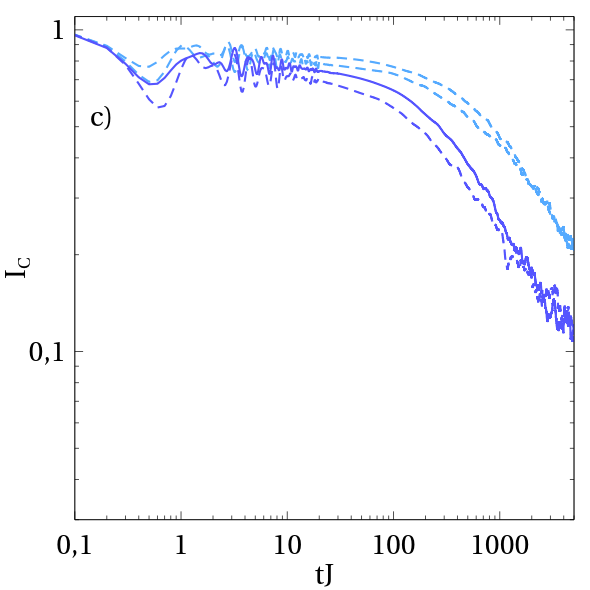}\includegraphics[scale=0.3]{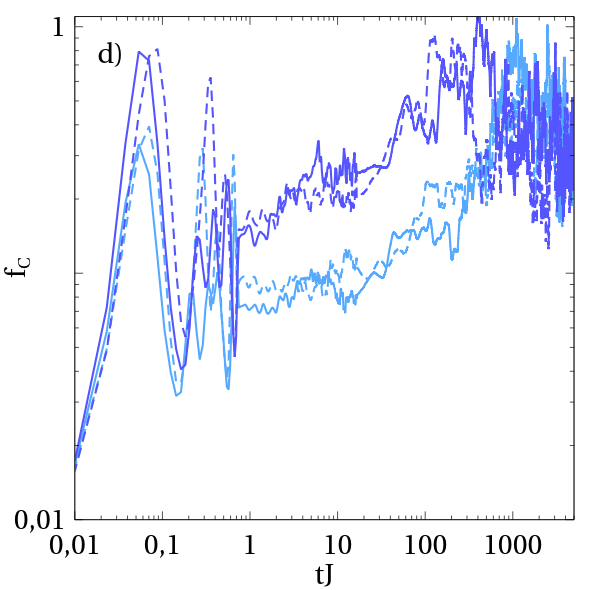}
\includegraphics[scale=0.3]{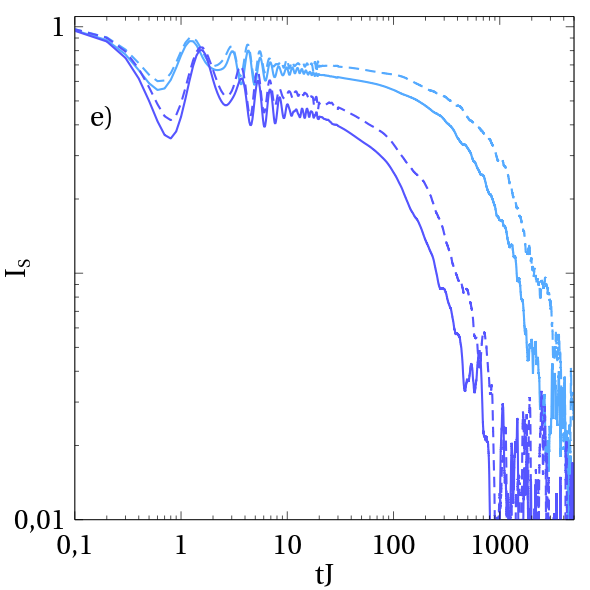}\includegraphics[scale=0.3]{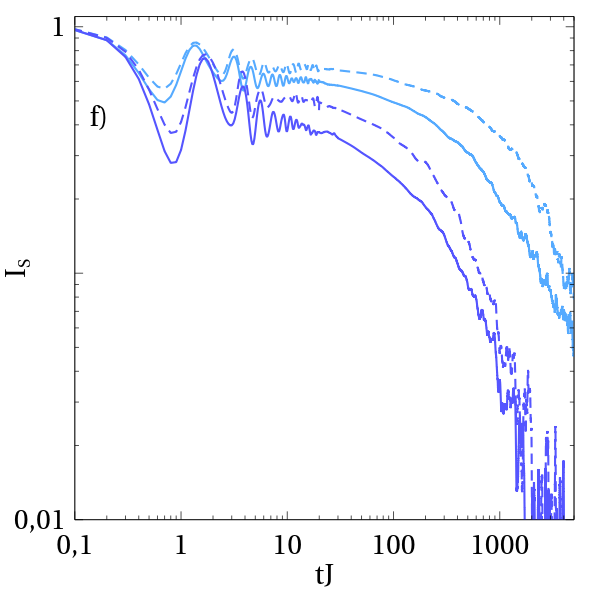}\includegraphics[scale=0.3]{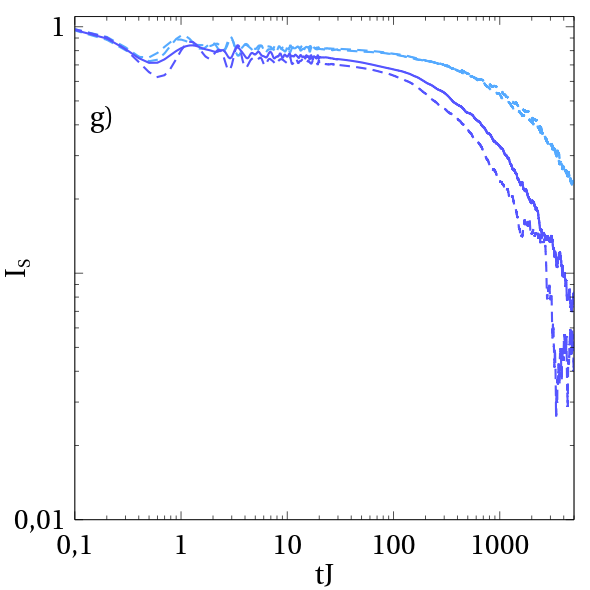}\includegraphics[scale=0.3]{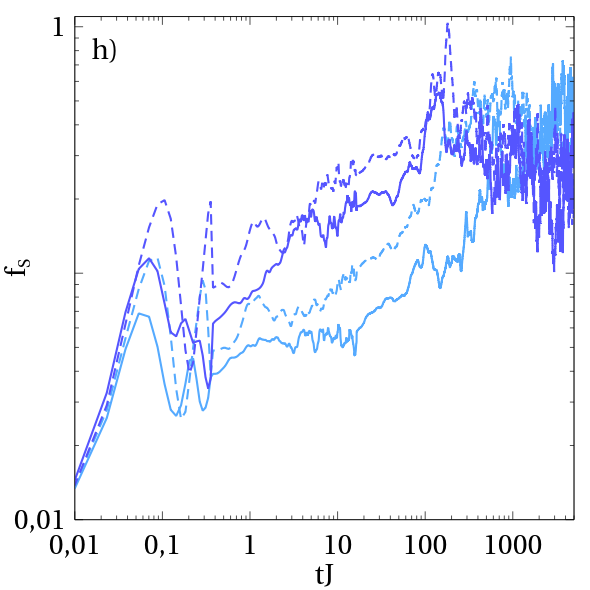}
\caption{Imbalance $I_{C}$, $I_{S}$ and QFI $f_{C}$, $f_{S}$ functions
for different lattice sizes: $4\times4$ - dashed lines, $6\times6$
- solid lines. In each figure the plots are made for the two values of tilt strength
$\Delta_{1}/J=4$ (dark blue), $\Delta_{2}/J=5$ (light blue). The first
row (Fig. a-d) represents the dynamics evaluated from the striped CDW initial condition,
while the second row (Fig. e-f) represents the dynamics evaluated from the striped SDW initial
condition. $U/J=1$ and about 100 fTWA trajectories where used for
a simulation of each line. \label{fig: finite-size effects}}
\end{figure*}

\begin{figure}
\includegraphics[scale=0.29]{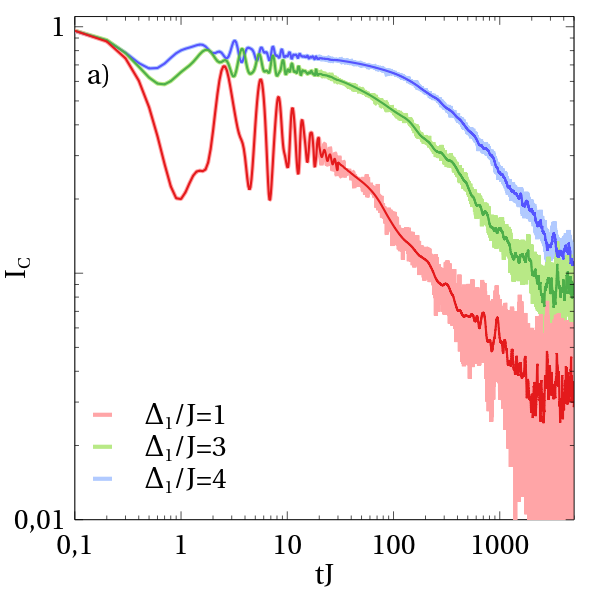}\includegraphics[scale=0.29]{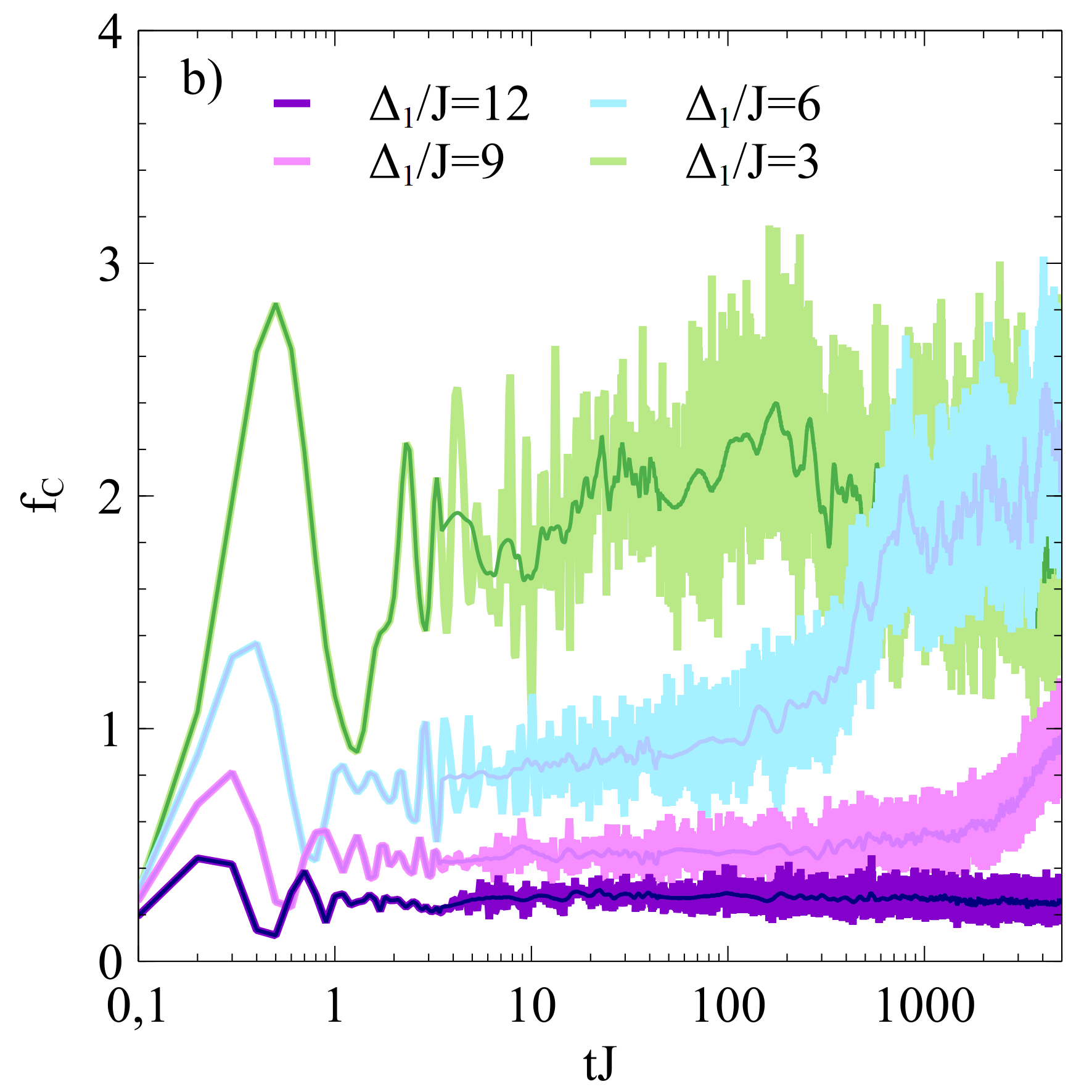}
\caption{Charge imbalance $I_{C}$ (Fig. a) and QFI $f_{C}$ (Fig. b) for different
values of the linear potential $\Delta_{1}/J$. Solid lines inside envelopes
representing original data are obtained by filtering high frequency
oscillation and fTWA noise. The data presented in (a) and (b) correspond
to Fig. \ref{fig: imbalance 2d} c and Fig.~\ref{fig: QFI 2d}~a for
chosen values of $\Delta_{1}/J$. \label{fig: filtering}}
\end{figure}

In this work we show that for a certain range of parameters, the fTWA method
can efficiently simulate quantum-many body dynamics for the tilted
Hubbard model. This is the case when a harmonic and spin-dependent linear potentials are imposed. Interestingly we observe also that this improvement appears
for higher-order correlation functions like QFI suggesting that this
strictly non-meanfield result is very efficiently described by the
quantum fluctuation included in fTWA.

These results enable us to discuss the many-body dynamics of the disorder-free
two-dimensional square lattice. We show that quantum evolution of
charge and spin degrees of freedom exhibits subdiffusive behavior
which is similar to that of disorder systems \citep{BarLev2016,PhysRevA.102.033338}
(however, in two-dimensional disordered systems a behavior somewhat
faster than a power law one can be expected due to rare regions \citep{PhysRevB.93.134206,Luitz2017,PhysRevLett.125.155701,P_pperl_2021}).
Moreover, disorder-like dynamical behavior is also recovered for QFI
which show a logarithmic-like growth \citep{PhysRevB.99.241114,Guo2020,PhysRevA.102.033338}.
Next focusing our study on the on-site density dynamics, we show that
the harmonic potential induces lattice locations at which the ergodic
or non-ergodic type of behavior is observed. This result complements
the recent studies in one dimension in which phase separation of ergodic
and non-ergodic regions has been observed \citep{Yao2021,PhysRevResearch.2.032039,PhysRevB.102.104203,Morong2021-mh}.

It is also worth pointing out that the spin dependence of the linear potential, controlled in our simulations by a parameter $A$,
 was similar to that of the recent experimental
work in Ref. \citep{Scherg2021}. This suggests that the fTWA method
can become an efficient tool for the theoretical prediction of real
experimental data for larger system sizes.

In future studies it will be interesting to investigate other types
of initial conditions like domain walls in two dimensions \citep{PhysRevB.105.134204}
or other types of tilts that modify the lattice directions for which
potential changes can be small \citep{vanNieuwenburg2019}. Moreover
the harmonic potential strength analyzed in this work can induce anomalously
slow dynamics in different parts of the lattice locations therefore
it will be also interesting to look at the dynamics locally
and test the local effective potential theory in the semiclassical
picture \citep{PhysRevB.104.014201,PhysRevResearch.2.032039,PhysRevB.102.104203,Yao2021}.

\section{Acknowledgments}

We would like to thank Marcin Mierzejewski for valuable discussions. A.S.S. acknowledges the funding from the Polish Ministry of Science and Higher Education through a ''Mobilno\'{s}\'{c} Plus'' program nr 1651/MOB/V/2017/0. Numerical studies in this work have been carried out using resources provided by the Wroclaw Centre for Networking and
Supercomputing  \footnote{http://wcss.pl}, Grant No. 551, and in part also by PL-Grid Infrastructure.

\section{Appendix: Finite-size effects \label{sec:Appendix:-Finite-size-effects}}

In Fig. \ref{fig: finite-size effects} we present the finite-size effects
for the imbalance function and QFI. Within the considered system sizes, we
only see qualitative difference in these effects for the charge channel
without an imposed harmonic potential (Fig. \ref{fig: finite-size effects}
a, b). Interestingly, in the limit of disorder-like behavior observed 
for the two-dimensional system, Fig. \ref{fig: finite-size effects}
c, d, g, h, the finite-size effects seem unimportant in the presented discussion.

\section{High frequency oscillations and noise}

The dynamics presented in Figs. \ref{fig: filtering} was filtered from
high frequency oscillations coming from inherit quantum dynamics on the tilted
lattice and from spurious fTWA noise coming from sampling of the initial
Wigner function. The spurious fTWA noise can be removed taking more
fTWA trajectories, however, the application of filtering is less numerically
costly than the simulation of more trajectories. We checked that addition of more
fTWA trajectories does not change the filtered signal. To filter the obtained data a Kaiser window was used.
An examplary original signal and its form after filtering are presented
for $I_{C}$ and $f_{C}$ in Fig. \ref{fig: filtering}.

\bibliographystyle{apsrev4-1}
\bibliography{library}

\end{document}